\documentclass[aps,prd,amsmath,amssymb,superscriptaddress,notitlepage,12pt,longbibliography]{revtex4-1}
\usepackage[utf8]{inputenc}
\usepackage{graphicx}
\usepackage{hyperref}
\usepackage{mathrsfs}

\begin{document}

\title{Anti-de Sitter geon families}
\author{Gyula Fodor}
\affiliation{Wigner Research Centre for Physics, RMKI, 1525 Budapest 114, P.O.~Box 49, Hungary}
\author{P\'eter Forg\'acs}
\affiliation{Wigner Research Centre for Physics, RMKI, 1525 Budapest 114, P.O.~Box 49, Hungary}
\affiliation{LMPT, CNRS-UMR 6083, Universit\'e de Tours, Parc de Grandmont, 37200 Tours, France}

\date{\today}

\begin{abstract}
A detailed perturbative construction of globally regular, asymptotically anti-de Sitter (AdS) time-periodic solutions of Einstein's equations with a negative cosmological constant (AdS geons) is presented.
Starting with the most general superposition of the $l=2$ even parity (scalar) eigenmodes of AdS at linear order, it is shown that at the fifth order in perturbation theory one obtains five one-parameter geon families, two of which have a helical Killing vector, one with axial symmetry, and two others without continuous symmetries. The details and some subtle aspects of the perturbative expansions are also presented.
\end{abstract}

\maketitle
\section{Introduction} \label{sec:intro}

The gravitational dynamics of asymptotically anti de Sitter (AAdS) spacetimes has attracted considerable interest in the past few years, stimulated to a large extent by the AdS/CFT correspondence.
In the seminal paper \cite{BizonRostw2011}, the time evolution of a free, massless scalar field coupled to Einstein's gravity has been investigated in three spatial dimensions in AAdS spacetimes, with the result that from a large class of smooth initial data black holes form, indicating that AAdS is unstable against black hole formation.
The instability uncovered by Ref.\ \cite{BizonRostw2011} manifests itself by the concentration of more and more energy in the same spatial region, where the mechanism is usually referred to as weak turbulence.
A considerable amount of work has followed (see e.g., the reviews \cite{Bizon14,Craps16,Gregoire17PhD,Gregoire17}), and by now there is little doubt that AAdS spacetimes exhibit weakly turbulent-type instabilities leading to black hole formation for a large class of initial data.
Asymptotically AdS spacetimes possess a peculiar causality structure; they are not globally hyperbolic: i.e., there is no Cauchy hypersurface in them.
Therefore it is not sufficient to specify the initial data on a spacelike hypersurface to determine the time evolution
in AAdS spacetimes. Because of the presence of timelike conformal boundaries of AAdS spacetimes at (null and spatial) infinity suitable boundary conditions have to be imposed on the fields.

It is important to note that the peculiarities of AAdS spacetimes also allow for the existence of various spatially localized (``particlelike'') objects of finite mass in various field theories. The boundary conditions induced by the negative cosmological constant make possible the existence of a much larger class of particlelike solutions than in asymptotically Minkowskian or de Sitter spacetimes.

The central objects of interest of the present paper are spatially localized, time-periodic solutions of Einstein's equations in AAdS spacetimes, referred to as AdS breathers.
In non-AAdS spacetimes breather-type solutions exist only under very special circumstances, since in most field theories breather-type initial data would evolve in general to some radiating object (oscillons, oscillatons) because of the presence of a continuous spectrum.
Spherically symmetric AdS breathers occur in scalar theories, and it is important that some of them actually appear to be stable against collapsing to a black hole
\cite{Buchel12,Maliborski13,Fodor14,Green15,Deppe15b,Dimitrakopoulos15,Craps15,Fodor15,Kim15,Duarte16,Herdeiro16}. The existence of such stable breathers
indicate the presence of stability islands in asymptotically AdS spacetimes, likely forming sets of nonzero measure initial data\cite{Dias12b}.
The existence of known AdS breathers is intimately related to the existence of normalizable Fourier eigenmodes of the wave operator in AdS spacetimes for special values of the frequency (this phenomenon is absent in the asymptotically Minkowski or the deSitter case). A standard way to start the perturbative (or numerical) construction of AdS breathers is to deform such normalizable AdS eigenmodes nonlinearly in order to construct solutions of the full theory.

The central aim of the present paper is to present a detailed perturbative construction of AdS gravitational breathers in Einstein's gravity with a negative cosmological constant. Gravitational AdS breathers are referred to in the literature as ``AdS geons,'' nomenclature we shall also use in the following.
In contradistinction to spherically symmetric breathers, gravitational AdS geons have fewer symmetries, and therefore are more difficult to construct. The first example of an AdS geon with a helical Killing vector has been constructed in perturbation theory in Ref.\ \cite{Dias12a} and subsequent analytical and numerical works have considerably enlarged our knowledge on AdS geons and on their stability \cite{Dias12b,Dias16a,Rostwor2017a,Horowitz14,Gregoire2017}.
The trustworthiness of the perturbative geon construction has been greatly enhanced by the numerical results of Refs.\ \cite{Horowitz14,Gregoire2017}, leaving little doubt as to the existence of helically symmetric AdS geons, being globally regular solutions of Einstein's equations. Besides these helically symmetric geons possessing angular momentum, $J\ne0$, strong hints of the existence of axially symmetric geons with $J\equiv0$ have been produced by perturbation theory in Ref.\ \cite{Dias17a}. Such nonrotating geon solutions may seem somewhat counterintuitive, as they consist just of gravitational waves and one would expect that rotation is necessary to counteract gravitational attraction.

In this paper we carry out a complete analysis of AdS geons 
corresponding to the superposition of the simplest scalar perturbations with $l=2$ at linear order
with frequency $\omega=3$. 
Starting with the most general superposition at linear order (depending on ten parameters) we show that to fifth order in perturbation this reduces to five (inequivalent) one-parameter geon families. It turns out that one really has to push the perturbative approach to fifth order to establish the result, since at third order there is still a two-parameter family of geons satisfying the consistency conditions, which splits to two one-parameter families at fifth order.
We carry out a full fledged fifth order computation whereby we recover and improve upon the known helically symmetric and axially symmetric geon families. We have tried to include all necessary technical details in Appendixes \ref{app:asymptotics}-\ref{apptimeper} to make our work more useful and reproducible, including  some subtle but essential points in order to be able to solve the fifth order consistency conditions.
It is important that we have discovered two new geon families with zero angular momentum, which have no obvious continuous symmetries.
We present the most important physical properties of the geon families we have found, consisting of the relations between their frequencies, masses and angular momenta.

The paper is organized in the following way. In Sec.~\ref{secexpads} we present the basic formalism for the nonlinear perturbative expansion of time-periodic asymptotically AdS geon solutions. The specification of the conformal boundary conditions, together with the associated conserved quantities, is presented in Appendix \ref{app:asymptotics}. The globally regular localized linear order solutions are discussed in Sec.~\ref{seclingeon}. The used scalar- and vector-type real spherical harmonic functions and their properties are detailed in Appendix \ref{appsphharm}. In Sec.~\ref{seclofrgeonlin} we introduce the class of geon solutions considered in this paper, namely those that reduce to a combination of the $\omega=3$ frequency linear modes in the small amplitude limit. The effect of spatial rotations on the $l=2$ modes is detailed in Appendix \ref{app:rotations}. In Sec.~\ref{sec-highord} we present the most important points of the nonlinear perturbation formalism on AdS spacetime. The definition of the perturbative quantities are motivated by the Kodama-Ishibashi-Seto gauge invariant formalism, which is discussed in Appendix \ref{apphighordg}. All spherical harmonic components of the metric perturbations are generated by scalar functions that satisfy second order inhomogeneous wave equations. The method of the introduction of these generating functions, and the calculation of the metric from them is presented for vector- and scalar-type perturbations in Appendixes \ref{app:vectpert} and \ref{app:scalpert}, respectively. Time-periodic solutions of the scalar wave equations are constructed in Appendix \ref{apptimeper}. The role of the arising resonance conditions and of the appearing free parameters that correspond to amplitudes of regular homogeneous solutions is explained in Sec.~\ref{sec-highord}. In Sec.~\ref{sechighordlowfr} the previously introduced $\omega=3$ solutions are considered, and it is shown that there are five one-parameter families that satisfy the fifth order consistency conditions. Section \ref{secfivefamily} contains the fifth order expansion results for these five families of solutions, listing the relations between the frequencies, masses and angular momentums. There is also a conclusions section at the end of the main part of the manuscript.

\section{Expansion of Anti-de Sitter geons} \label{secexpads}

We consider $(3+1)$-dimensional vacuum Einstein equations, $G_{\mu\nu}+\Lambda g_{\mu\nu}=0$, where the cosmological constant $\Lambda$ is negative, and it is related to the length scale $L$ by
\begin{equation}
 L^2=-\frac{3}{\Lambda} \ .
\end{equation}
We look for solutions of Einstein's equations perturbatively, assuming that the solution depends on a small parameter, $\varepsilon$, in terms of which the metric tensor can be expanded in power series as
\begin{equation}
 g_{\mu\nu}=\sum_{k=0}^\infty \varepsilon^k g_{\mu\nu}^{(k)} \ . \label{gmunuexp1}
\end{equation}
The tensor $g_{\mu\nu}^{(k)}$ represents $k$th order nonlinear perturbations of the background spacetime $g_{\mu\nu}^{(0)}$. For our calculations we use coordinates $x^\mu=(t,x,\theta,\phi)$ and set the zeroth order term $g_{\mu\nu}^{(0)}$ to the anti-de Sitter metric in the form
\begin{equation}
 \mathrm{d}s^2_0=\frac{L^2}{\cos^2 x}\left(-\mathrm{d}t^{\,2}+\mathrm{d}x^2
 +\sin^2 x \, \mathrm{d}\bar\Omega^2\right) \ , \label{sphttx}
\end{equation}
where $\mathrm{d}\bar\Omega^2=\mathrm{d}\theta^2+\sin^2\theta\mathrm{d}\phi^2$ is the standard metric on the 2-sphere. The naturally defined radius function is
\begin{equation}
r=L\tan x \ .
\end{equation}
Schwarzschild-type coordinates can be obtained by using $r$ as a radial coordinate, and introducing a rescaled time coordinate $\bar t=Lt$. The physical time coordinate is $\bar t$, because it agrees with the proper time of the AdS background at the center.

We choose an $\varepsilon$ independent conformal factor
\begin{equation}
 \Omega=\frac{\cos x}{L} \ . \label{eqconffact}
\end{equation}
It follows from \eqref{sphttx} that the conformally transformed metric obtained from the background solution, $\tilde g_{\mu\nu}^{(0)}=\Omega^2 g_{\mu\nu}^{(0)}$, is regular at the surface $x=\pi/2$. This three-dimensional timelike hypersurface corresponds to conformal infinity and is denoted by $\mathscr{I}$. A natural approach would be to require the boundary condition $\displaystyle\lim_{x\to\pi/2}\left(\Omega^2 g_{\mu\nu}^{(k)}\right)=0$ for all $k\geq 1$, which would ensure that for the whole one-parameter family the conformally transformed metric, $\tilde g_{\mu\nu}=\Omega^2 g_{\mu\nu}$, would remain unchanged on $\mathscr{I}$; hence regular and, consequently, each solution in the family would be asymptotically anti-de Sitter. We give a more detailed description of the requirements that asymptotically AdS spacetimes have to satisfy in Appendix \ref{app:asymptotics}. With this simple choice of boundary conditions the time coordinate $t$ would remain the natural time coordinate that agrees asymptotically with the AdS time coordinate. To distinguish from the time coordinate that we will use throughout the paper we denote this $t$ by $\hat t$. For the determination of the oscillation frequencies measured by distant observers the Schwarzschild-type time coordinate $\bar t=L\hat t$ has to be used.

The $\bar\omega$ physical frequency of the geon solutions changes as their amplitude increases, so it is generally $\varepsilon$ dependent. To make the expansion calculations technically much simpler we use a time coordinate $t$ such that with respect to this $t$ the coordinate frequency $\omega$ of the geons remains constant. This can be achieved by requiring the boundary conditions
\begin{align}
 \lim_{x\to\frac{\pi}{2}}\left(\Omega^2 g^{(k)}_{tt}\right)&=-\nu_k \ , \label{nukdefeq}\\
 \lim_{x\to\frac{\pi}{2}}\left(\Omega^2 g^{(k)}_{\mu\nu}\right)&=0 \quad \text{for} \  \mu\neq t \  \text{or} \ \nu \neq t \  \label{eqgmunulim},
\end{align}
where $\nu_k$ are constants, independent of the angular coordinates. Because of the $\varepsilon\to-\varepsilon$ symmetry of the system, $\nu_k$ are nonzero only for even $k$. Since $\nu_k$ are assumed to be constants, the limit of $\Omega^2 g^{(k)}_{tt}$ is zero for all spherical harmonic components, except for the $l=0$, $m=0$ spherically symmetric part. Then the asymptotic behavior of the $g_{tt}$ metric component is $\varepsilon$ dependent,
\begin{equation}
 \lim_{x\to\frac{\pi}{2}}\left(\Omega^2 g_{tt}\right)=-\nu
 \quad , \qquad \nu=1+\sum_{k=1}^\infty\varepsilon^k\nu_k \ . \label{eqnudef}
\end{equation}
It follows that the asymptotically AdS time coordinate is $\hat t=t\sqrt{\nu}$. The physical frequency $\bar\omega$ has to be calculated with respect to a time coordinate that asymptotically agrees with the Schwarzschild time coordinate $\bar t=L\hat t=tL\sqrt{\nu}$, and the relation between the two frequencies is
\begin{equation}
 \bar\omega=\frac{\omega}{L\sqrt{\nu}} \ . \label{eqomeganu}
\end{equation}

The requirements \eqref{nukdefeq} and \eqref{eqgmunulim} on the asymptotic form of the metric perturbations ensure that the resulting metric $g_{\mu\nu}$ will be asymptotically AdS. As it is described in more detail in Appendix \ref{app:asymptotics}, this follows from the fact that the metric generated at asymptotic infinity remains essentially the same as that of the AdS metric; in particular, it remains conformally flat.

\section{Linear order geons} \label{seclingeon}

The study of perturbations of spherically symmetric spacetimes was initiated by the seminal paper of Regge and Wheeler\cite{ReggeWheeler}, taking the $(3+1)$-dimensional Schwarzschild spacetime as the background. They have shown that odd parity (also called axial) perturbations are governed by a scalar function that satisfies a wave equation. It was demonstrated later by Zerilli\cite{ZerilliPRL,ZerilliPRD} that even parity (polar) perturbations are also generated by a single scalar function. Gauge invariant variables were first applied by Moncrief\cite{Moncrief1974}, and later by Gerlach and Sengupta\cite{GerlachSengupta1,GerlachSengupta2,GerlachSengupta3} for general four-dimensional spherically symmetric spacetimes. Arbitrary dimensional anti-de Sitter background was first considered by Mukohyama\cite{Mukohyama2000}. A gauge invariant formalism for general $(\hat m+\hat n)$-dimensional spherically symmetric background spacetimes, where $\hat n$ is the dimension of the symmetry spheres, was worked out in detail by Kodama, Ishibashi and Seto \cite{KodIshSet}. When $\hat m=2$, for arbitrary $\hat n$, a generalization of the Regge-Wheeler scalar exists. The generalization of the Zerilli function for the $\hat m=2$ case was presented by Kodama and Ishibashi in \cite{KodIsh2003}. When the dimension of the symmetry spheres is greater than two, in addition to even and odd parity perturbations there is a third type of perturbation. Since this can be expanded in terms of tensor spherical harmonic functions, it is called tensor-type perturbation. Tensor-type perturbations were considered first for cosmological problems, since the symmetry sphere of the background is three dimensional in that case \cite{Bardeen1980,KodamaSasaki}. We follow this more general terminology in this paper, calling even parity (polar) perturbations as scalar-type and odd parity (axial) perturbations as vector-type. A detailed description of the linear perturbation formalism of anti-de Sitter spacetime is given by Ishibashi and Wald in \cite{IshWal}, which was a very important reference during our work.

General solutions of the $(3+1)$-dimensional linearized Einstein equations around the AdS background \eqref{sphttx} can be uniquely decomposed into the sum of scalar- and vector-type perturbations. Both types of perturbations are further decomposed into spherical harmonic classes indexed by the integers $l$, and $m$ with $-l\leq m \leq l$. Perturbations with $l=0$ or $1$ are pure gauge modes at linear order, so we do not consider them at this stage, i.e.\ we shall assume $l\geq 2$.
Classes of perturbations belonging different $l$, $m$, and to scalar- and vector-types decouple in the linear equations, because of the rotational invariance of the background and the operators. For the linear case, in each class, the asymptotically AdS centrally regular perturbations are explicitly known\cite{IshWal}. They are linear combinations of time periodic solutions labeled by a non-negative integer $n$. The time phase of the solutions is arbitrary, which we take into account by allowing two contributions with independent amplitudes, one with cosine and the other with sine time dependence.

Regular scalar-type perturbations exist only for frequencies
\begin{equation}
\omega^{(S)}_{ln}=l+1+2n \ ,  \label{eqscalfreq1}
\end{equation}
and they are generated by the functions 
\begin{align}
\Phi^{(Sc)}_{lmn}&=\alpha^{(Sc)}_{lmn}\cos\left(\omega^{(S)}_{ln} t\right)p^{(S)}_{ln} \ , \label{scallinpert1} \\
\Phi^{(Ss)}_{lmn}&=\alpha^{(Ss)}_{lmn}\sin\left(\omega^{(S)}_{ln} t\right)p^{(S)}_{ln} \ , \label{scallinpert2}
\end{align}
where $p^{(S)}_{ln}$ are functions of $x$, and $\alpha^{(S\sigma)}_{lmn}$ are arbitrary constant amplitudes for each choice of $l$, $m$, $n$ and $\sigma=(s\text{ or }c)$. The explicit expression for $p^{(S)}_{ln}$ is given by \eqref{pslnsol} in Appendix \ref{apptimeper}, where the nonlinear formalism is discussed in detail. In this section the important points are the time dependence of $\Phi^{(S\sigma)}_{lmn}$ and the number of independent constants $\alpha^{(S\sigma)}_{lmn}$. For fixed $l$, $n$ and $\sigma$ there are $2l+1$ independent modes for each $m$ in the range $|m|\leq l$, and the generated perturbations differ only in their angular behavior.

Regular vector-type perturbations exist only for frequencies
\begin{equation}
\omega^{(V)}_{ln}=l+2+2n \label{eqvectfreq1}
\end{equation}
and are generated by the functions
\begin{align}
\Phi^{(Vc)}_{lmn}&=\alpha^{(Vc)}_{lmn}\cos\left(\omega^{(V)}_{ln} t\right)p^{(V)}_{ln} \ ,
\label{vectlinpert1} \\
\Phi^{(Vs)}_{lmn}&=\alpha^{(Vs)}_{lmn}\sin\left(\omega^{(V)}_{ln} t\right)p^{(V)}_{ln} \ ,
\label{vectlinpert2}
\end{align}
where $p^{(V)}_{ln}$ are given in \eqref{eqpvlndef}. The integer $n$ gives the number of radial nodes (zero crossings) of the generating functions. 

Each $\Phi^{(\Sigma\sigma)}_{lmn}$ function, where $\Sigma=(S\text{ or }V)$, generates a part of the $g_{\mu\nu}^{(1)}$ linear metric perturbation that belongs to the $l,m$ scalar or vector spherical harmonic class. The general scalar or vector $l,m$ class perturbation is the linear combination of these for all $n\geq 0$. The details of the differential map that gives the metric tensor components from the scalars $\Phi^{(\Sigma\sigma)}_{lmn}$ are given in Appendix \ref{app:vectpert} for vector- and Appendix \ref{app:scalpert} for scalar-type perturbations, where the formalism for general order in $\varepsilon$ is discussed in detail. The important point here is that all the amplitude constants $\alpha^{(\Sigma\sigma)}_{lmn}$ can be chosen independently, and all the contributions give time-periodic perturbations with integer frequencies. Any combination of these will be periodic at least with frequency $\omega=1$. This shows that if we consider the formalism only to first order, we have an infinite-parameter family of linear geon solutions. However, only a small part of these solutions will correspond to truly periodic solutions of the nonlinear system. In fact, in all cases that we have studied, only single-parameter families survive the consistency conditions that arise at higher order in the expansion formalism.

A natural first task is to consider cases when only one of the constants $\alpha^{(\Sigma\sigma)}_{lmn}$ is nonzero. This approach was followed in \cite{Dias12a,Horowitz14,Dias16a,Dias17a}, concentrating on helically or axially symmetric solutions. It was shown in these papers that only some exceptional single linear modes survive to higher orders. However, as it was pointed out in \cite{Rostwor2017a,Rostwor2017b,Rostwor2017c}, modes with identical frequency should be combined even at linear order. For the cases investigated in these papers, the same frequency linear combinations have been observed to give as many one-parameter families of nonlinear solutions as the multiplicity of the given frequency. In \cite{Gregoire2017} three one-parameter families of helically symmetric AdS geon solutions were presented. To first order in the expansion each of them reduces to a nontrivial linear combination of three modes, a vector mode, a scalar mode with one radial node, and a scalar mode without radial nodes, all three with frequency $\omega=5$.

\section{Lowest frequency geons} \label{seclofrgeonlin}

One of the main aims of the present paper is to present the complete classification of those geon families that in the small amplitude limit oscillate only with the lowest possible frequency. It follows from \eqref{eqscalfreq1} that in the linear order, the expansion of these solutions includes only scalar modes with $l=2$ and $n=0$, and the frequency is $\omega=3$. As \eqref{eqvectfreq1} shows, the frequency of the vector modes is at least $\omega=4$, so for these types of solutions vector modes can be generated only at higher orders in the $\varepsilon$ expansion. For the linear scalar modes there are five integer values in the $-2\leq m\leq 2$ interval, and since for all of these the time dependence can be sine or cosine, altogether there are ten freely specifiable constants, $\alpha^{(S\sigma)}_{2m0}$. We consider two solutions equivalent if one can be transformed into the other by spatial rotations and by a constant time shift. A general spatial rotation can be specified by three Euler angles $\alpha$, $\beta$ and $\gamma$. One can expect to be able to make zero three coefficients from the ten numbers $\alpha^{(S\sigma)}_{2m0}$ by appropriate choice of the Euler angles. We use the rotational freedom to set $\alpha^{(Sc)}_{210}=\alpha^{(Sc)}_{2-10}=\alpha^{(Sc)}_{2-20}=0$, keeping only the $m=2$ and $m=0$ components nonzero from the constants $\alpha^{(Sc)}_{2m0}$ corresponding to the cosine time dependence. It is shown in Appendix \ref{app:rotations} that this can always be achieved. The five coefficients $\alpha^{(Ss)}_{2m0}$ with the sine time dependence can be arbitrary at this stage, so we still have seven parameters. Further restrictions on the parameters, from the nonlinear nature of the problem, will only come at $\varepsilon^3$ and $\varepsilon^5$ orders in the small-amplitude expansion procedure.

\section{Higher order expansion in general} \label{sec-highord}

We do not attempt here to work out a general formalism to write down explicit expressions for the $\varepsilon^k$ order components of the Einstein equations. See, for example, \cite{Rostwor2017b} for such results. Nonlinear perturbations of Einstein equations can be efficiently calculated to very high orders by some algebraic manipulation software, such as Maple, Mathematica or SageMath. If the covariant components of the metric are known up to the order $K$,
\begin{equation}
 g_{\mu\nu}=\sum_{k=0}^K \varepsilon^k g_{\mu\nu}^{(k)} \ , \label{eqgmunufinite}
\end{equation}
then the contravariant components of the metric can be calculated by the chosen algebraic manipulation program, and can be expanded up to order $K$ in $\varepsilon$. After this step, one only has to do multiplications and differentiations to compute the components of Christoffel symbols, the Riemann curvature tensor, and the Einstein equations $G_{\mu\nu}+\Lambda g_{\mu\nu}=0$. In each case when one has to multiply two expressions, in order to save memory and time, it is reasonable to calculate separately the $\varepsilon^k$ components of the product for $k\leq K$ from the coefficients of the two terms, in order to ensure that higher than $K$ order components are never calculated and stored. Although it is relatively easy to calculate the high order components of the Einstein equations by computer algebra, it is not possible to solve them without understanding their structure. At each order $k$ the equations contain linear terms in the unknown variables $g_{\mu\nu}^{(k)}$, which we discuss in the following paragraphs and in Appendixes \ref{app:vectpert} and \ref{app:scalpert} in detail, and also contain nonlinear source terms determined by lower order perturbations, which we obtain by computer algebra.

We $2+2$ decompose the background AdS metric according to its spherical symmetry\cite{GerlachSengupta1,KodIshSet},
\begin{equation}
 ds_0^2=\hat g_{ab}(y)dy^a dy^b+r^2(y)d\bar\Omega^2 \ ,  \label{eqbackgrm}
\end{equation}
where $y^a\equiv(y^1,y^2)$ are the coordinates in the time-radius plane (constant angles), and $d\bar\Omega^2=\gamma_{ij}dz^idz^j$ represents the metric of the unit 2-sphere $\mathcal{S}^2$. On the time-radius plane the two-dimensional metric induced by the AdS metric is denoted by $\hat g_{ab}$, and the corresponding derivative operator by $\hat\nabla_a$. We use standard spherical coordinates $z^i\equiv(z^3,z^4)=(\theta,\phi)$, and on the time-radius plane we use coordinates $y^a=(t,x)$. 

Scalar functions on the two-dimensional sphere can be expanded in terms of real scalar spherical harmonics $\mathbb{S}_{lm}$, where $l$ and $m$ are integers satisfying $l\geq 0$ and $|m|\leq l$. The definition of the $\mathbb{S}_{lm}$ we use is given in Appendix \ref{appsphharm}. One-form fields and symmetric tensors can be uniquely decomposed into scalar-type and vector-type parts. The vector-type parts can be decomposed in terms of the vector spherical harmonics $\mathbb{V}_{(lm)i}$. A detailed description of the decomposition procedure and its practical application is given at the end of Appendix \ref{appsphharm}.

Higher order perturbations of four-dimensional AdS spacetime can be expanded as sums of scalar- and vector-type perturbations. At order  $\varepsilon^k$, perturbations in different classes (scalar and vector) and with different $l$ and $m$ are only coupled to each other through nonlinear source terms, which are completely determined by lower than $k$ order perturbations. We give a concise description of the nonlinear gauge-invariant perturbation formalism in Appendix \ref{apphighordg}. Since we are working with a concrete natural gauge choice at each order in the formalism, the deep understanding of the gauge invariant formalism is not necessary for the actual calculations.

At each order in the $\varepsilon$ expansion one can make a gauge choice and proceed order by order with the perturbation formalism. We choose a gauge in which the metric perturbation variables are very closely related to the Kodama-Ishibashi-Seto gauge-invariant variables\cite{KodIshSet}. This choice corresponds to the Regge-Wheeler gauge in the literature. Scalar- and vector-type perturbations must be treated separately. Perturbations with spherical harmonic index $l=0$ and $l=1$ also require special treatment.

We proceed order by order with the perturbation formalism. We consider $\varepsilon^k$ order perturbations, assuming that all the lower order perturbations are already calculated and fixed. If the construction is ready up to $\varepsilon^{k-1}$ order, then we can calculate the nonlinear source terms in the $\varepsilon^k$ order Einstein equations by some algebraic manipulation software and decompose the source terms into $(l,m)$ scalar and vector harmonic components using the procedure described in Appendix \ref{appsphharm}. There will be only a finite number of nonzero source term components at each order. We can solve each scalar and vector $(l,m)$ component equation separately for the $\varepsilon^k$ order perturbation quantities. In most cases, but not always, it is enough to consider only those components that have a nonzero nonlinear source term and take the trivial zero solution for the others. The calculations, especially the necessary integrations in the $x$ variable, are becoming more and more involved technically as the order of the expansion increases, but in many cases they can be managed up to order $\varepsilon^6$ by an algebraic manipulation software.

The formalism for vector-type components is somewhat simpler than that of the scalar-type. Because of the complicated technical details, vector-type perturbations are discussed in Appendix \ref{app:vectpert} and scalar-type perturbations in Appendix \ref{app:scalpert}. The equations determining the $k$th order perturbation of the metric for the spherical harmonic index $l=0$ or $l=1$ can always be solved directly, for both scalar- and vector-types, as it is demonstrated in Appendixes \ref{app:vectpert} and \ref{app:scalpert}. If there are no source terms arising from lower than $k$ orders, then in the $l\leq 1$ case only the trivial zero solution remains.

When $l\geq 2$, for each choice of $(l,m)$, the vector-type perturbation of the metric tensor is generated by a scalar function $\Phi_V$, and similarly, class $(l,m)$ scalar-type perturbations are generated by another scalar $\Phi_S$. To make notation shorter, the indices $l$ and $m$ are dropped from $\Phi_V$ and $\Phi_S$. As it is shown in Appendixes \ref{app:vectpert} and \ref{app:scalpert}, both of these functions satisfy a differential equation having the form
\begin{equation}
 -\frac{\partial^2\Phi}{\partial t^2}+\frac{\partial^2\Phi}{\partial x^2}
 -\frac{l(l+1)}{\sin^2 x}\Phi+\frac{\Phi^{(0)}}{\sin^2 x}=0 \ . \label{mastereqgen}
\end{equation}
This equation is the master equation describing all $l\geq 2$ scalar- and vector-type $\varepsilon^k$ order perturbations. Since we are interested in geon configurations, we are looking for time-periodic solutions of \eqref{mastereqgen}. The boundary conditions at infinity are different in the vector and scalar cases. 

The homogeneous part of \eqref{mastereqgen} is the same in all cases, but the inhomogeneous source term $\Phi^{(0)}/\sin^2 x$ arising from lower order perturbations is generally different for each vector or scalar $(l,m)$ component. The actual form of $\Phi^{(0)}$ is obtained by an algebraic manipulation software in each case. Generally, the source term is the sum of a finite number of time-periodic (or static) terms,
\begin{equation}
 \Phi^{(0)}=\sum_{\alpha=1}^{\alpha_{\mathrm{max}}}\left[
 p^{(0c)}_\alpha\cos(\omega_\alpha t)+p^{(0s)}_\alpha\sin(\omega_\alpha t)\right] , \label{phi0expeq}
\end{equation}
where $\omega_\alpha$ are non-negative integers, and the functions $p^{(0\sigma)}_\alpha$, for $\sigma=c$ or $s$, depend only on the radial coordinate $x$. The natural way to solve Eq.~\eqref{mastereqgen} is to solve the inhomogeneous equation separately for each individual term in \eqref{phi0expeq}, and then to add to the sum of these the general time-periodic solution of the homogeneous equation. This means that we have to solve ordinary differential equations for a function $p$ depending on $x$ having the form
\begin{equation}
\frac{{\rm d}^2p}{{\rm d} x^2}-\frac{l(l+1)}{\sin^2 x}p+\omega_p^2 p+\frac{p^{(0)}}{\sin^2 x}=0
\ . \label{eqpinhom}
\end{equation}
The frequency of each particular solution is given by the frequency of the corresponding source term, so in that case $\omega_p=\omega_\alpha$, and $p^{(0)}$ is one of $p^{(0c)}_\alpha$ or $p^{(0s)}_\alpha$. If one is looking for the solution of the homogeneous problem, then $p^{(0)}=0$, and $\omega_p$ is yet undetermined. The detailed description of the solution procedure of \eqref{eqpinhom} is presented in Appendix \ref{apptimeper}. The solution for the generating function has the form
\begin{equation}
 \Phi=\sum_{\beta=1}^{\beta_{\mathrm{max}}}\left[
 p^{(c)}_\beta\cos(\omega_\beta t)+p^{(s)}_\beta\sin(\omega_\beta t)\right] , \label{phiexpres}
\end{equation}
where $p^{(\sigma)}_\beta$ come from various solutions $p$ of \eqref{eqpinhom}, and the frequencies $\omega_\beta$ include all $\omega_\alpha$ from \eqref{phi0expeq} and some of the resonant frequencies of the homogeneous equation.

For a given $(l,m)$ vector- or scalar-type component let us find the particular solution corresponding to the part of the source term in \eqref{mastereqgen} with frequency $\omega_\alpha$. It follows from the results in Appendix \ref{apptimeper} that there are two cases to consider, resonant or nonresonant. The source term frequency $\omega_\alpha$ is resonant if for some non-negative integer value of $n$ it is equal to $\omega^{(S)}_{ln}=l+1+2n$ in the scalar-type case, or equal to $\omega^{(V)}_{ln}=l+2+2n$ in the vector-type case. For any nonresonant source term a unique centrally regular asymptotically AdS time-periodic particular solution of the master equation can always be obtained. (However, some integrals in $x$ may be extremely hard to perform.)

For a resonant source term centrally regular asymptotically AdS time-periodic solutions exist only if the following crucial consistency condition holds:
\begin{equation}
 \int_0^{\frac{\pi}{2}}\frac{p_1 p^{(0)}}{\sin^2 x}{\rm d}x=0 \ , \label{eqrescond}
\end{equation}
where $p^{(0)}$ is the radial part of the source term, and $p_1$ is the regular asymptotically AdS solution of the homogeneous problem, denoted by $p^{(S)}_{ln}$ or $p^{(V)}_{ln}$ in Appendix \ref{apptimeper}. 
To satisfy the consistency condition \eqref{eqrescond} one often has to include certain homogeneous solutions of the lower order perturbation equations whose amplitudes will then be determined (and be different from zero) by \eqref{eqrescond}.

Source terms with resonant frequencies will not only provide relations between the yet unspecified constants but also provide new constants. For these resonant frequencies the homogeneous equation has a solution that is behaving well both at the center and at infinity, and we can add this solution with an unspecified amplitude $c_p$ to the particular solutions. If the source term is nonzero, there is no reason to prefer the value $c_p=0$. What we observe at the actual calculations, is that each resonant source term at order $\varepsilon^k$ provides a new constant, and at order $\varepsilon^{k+2}$ we get constraints that restrict their values.

The regular time-periodic general solution of the homogeneous part of Eq.~\eqref{mastereqgen} contains infinitely many unspecified constants, exactly the same way as the linear problem discussed in Sec.~\ref{seclingeon}. For each non-negative integer $n$ there are solutions with frequency given by \eqref{eqscalfreq1} or \eqref{eqvectfreq1}, and the amplitude of each of them is arbitrary. At each order in $\varepsilon$ only one of these new constants can be canceled by the ambiguity in how the various states (or their corresponding initial data) are labeled by the parameter $\varepsilon$. We have only one freedom in determining the frequency change at each order in the expansion. We do not consider quasiperiodic solutions having modes with unrelated frequencies in this paper. The inclusion of the other constants generally correspond to physically different solutions. In the majority of cases they correspond to initial data that lead to non-time-periodic evolution. In our procedure, when considering time-periodic solutions only, the value of these constants will be restricted by consistency conditions at $2$ orders higher in $\varepsilon$. It is clearly impossible to treat too many unspecified constants even with the most modern algebraic manipulation software. A reasonable simplifying assumption is to set to zero the amplitude of all homogeneous modes for which there is no inhomogeneous source term in \eqref{mastereqgen} with the same frequency. Surprisingly, it turns out that we will have to make a few exceptions to this rule. At $\varepsilon^5$ order some modes give consistency conditions that can be solved only if we allow the same modes to appear with some nonzero amplitude already at $\varepsilon^3$ order, even if at that order there are still no source terms with their frequency.

\section{Higher order expansion of the lowest frequency geons} \label{sechighordlowfr}

Let us now return to the geon configurations considered in Sec.~\ref{seclofrgeonlin}, which in the small amplitude limit only have the $\omega=3$ frequency components. For the time being we keep all ten freely specifiable constants $\alpha^{(S\sigma)}_{2m0}$, which determine these solutions to linear order (here $\sigma=c$ or $s$, and $|m|\leq 2$). To ease notation we denote them simply by $\alpha_{\sigma m}\equiv\alpha^{(S\sigma)}_{2m0}$ from now on.

\subsection{Second order}

Proceeding to second order in $\varepsilon$, the nonlinear source terms will have scalar-type components with $l=0,2,4$, and vector-type components with $l=1,3$, with all possible $|m|\leq l$ belonging to these. The vector components are time independent, but in the scalar components, besides the static part, there are terms with $\cos(6t)$ and $\sin(6t)$ time dependence. Since $\omega=6$ cannot be written as $l+1+2n$ for $l=0,2,4$, it is not a resonant frequency, and so there are no consistency conditions to satisfy at second order in $\varepsilon$. We do not either introduce at this level any new constants that would give the amplitude of homogeneous solutions, because they are not motivated by the presence of corresponding source terms. At the $l=m=0$ component a new unspecified constant, $\nu_2$, must be introduced, according to \eqref{nukdefeq}, \eqref{eqhttnubar} and \eqref{eqnubarnuk}. Later, when its value will be known, it will determine the second order change of the oscillation frequency of the geon. Having the metric perturbation to second order, one can already calculate the leading order behavior of the mass and the angular momentum by the method described at the end of Appendix \ref{app:asymptotics},
\begin{align}
 M&=\frac{135}{512L^3}\varepsilon^2\sum_{m=-2}^{2}\left(\alpha_{cm}^2+\alpha_{sm}^2\right)
  , \label{eqmasstenpar} \\
 J_x&=\frac{45}{256L^2}\varepsilon^2\left(
  \sqrt{3}\alpha_{s0}\alpha_{c-1}-\sqrt{3}\alpha_{c0}\alpha_{s-1}
  +\alpha_{c-2}\alpha_{s1}+\alpha_{c-1}\alpha_{s2}
  -\alpha_{c1}\alpha_{s-2}-\alpha_{c2}\alpha_{s-1}\right) , \label{eqordtwojx}\\
 J_y&=\frac{45}{256L^2}\varepsilon^2\left(
  \sqrt{3}\alpha_{c0}\alpha_{s1}-\sqrt{3}\alpha_{s0}\alpha_{c1}
  -\alpha_{c-2}\alpha_{s-1}+\alpha_{c-1}\alpha_{s-2}
  +\alpha_{c1}\alpha_{s2}-\alpha_{c2}\alpha_{s1}\right) ,  \label{eqordtwojy} \\
 J_z&=\frac{45}{256L^2}\varepsilon^2\left(-2\alpha_{c-2}\alpha_{s2}-\alpha_{c-1}\alpha_{s1}
  +\alpha_{c1}\alpha_{s-1}+2\alpha_{c2}\alpha_{s-2}\right) . \label{eqordtwojz}
\end{align}

\subsection{Third order} \label{subsecthirdord}

At $\varepsilon^3$ order there are scalar source terms with $l=0,2,4,6$ and vector source terms with $l=1,3,5$, with time dependence $\cos(3t)$, $\sin(3t)$, $\cos(9t)$ and $\sin(9t)$. The $\omega=9$ frequency is resonant, but the consistency conditions \eqref{eqrescond} belonging to these terms are identically satisfied. The $\omega=3$ frequency is resonant only for the $l=2$ scalar modes, and for each value of $m=-2,-1,0,1,2$ we get two consistency conditions, one from the $\cos(3t)$ and another from the $\sin(3t)$ source term. Denoting these ten conditions by $C_{\sigma m}$, each of them contains one term that is a constant times $L^4\nu_2\alpha_{\sigma m}$, with the same $\sigma$ and $m$ as in their name $C_{\sigma m}$, and about $20$ other terms, which are third order homogeneous in the ten variables $\alpha_{\sigma m}$. Because of their length we do not present the detailed form of these conditions here.

We remind the reader that we consider two solutions equivalent if they can be transformed into each other by a spatial rotation and a time shift. To solve the conditions and to classify the solutions it is advantageous to employ the Euler rotation described in Sec.~\ref{seclofrgeonlin} and in Appendix \ref{app:rotations} to make $\alpha_{c1}=\alpha_{c-1}=\alpha_{c-2}=0$. The ten consistency conditions become somewhat shorter after this, but they are still cumbersome. It helps if we consider four cases separately, depending on which of $\alpha_{c2}$ and $\alpha_{c0}$ is zero.

For the first case we consider $\alpha_{c2}=\alpha_{c0}=0$. In this case only the $\alpha_{sm}$ components can be nonzero. Obviously, by a time translation we can make all $\alpha_{sm}=0$ and make $\alpha_{cm}$ equal to the previous values of $\alpha_{sm}$. We can also make more coefficients nonzero, setting $\alpha_{cm}/\alpha_{sm}$ as the same value for all $m$ with nonzero $\alpha_{sm}$. We do not consider these as different solutions. For this first case all ten conditions give the same relation,
\begin{equation}
 \nu_2=\frac{1221}{3584L^4\pi}\left(\alpha_{s-2}^2+\alpha_{s-1}^2+\alpha_{s0}^2
 +\alpha_{s1}^2+\alpha_{s2}^2\right) ,
\end{equation}
which determines the second order change in the oscillation frequency. The absence of further conditions shows that to third order in $\varepsilon$ all linear solutions with $\alpha_{cm}=0$ and arbitrary $\alpha_{sm}$ can be extended to valid nonlinear solutions. From \eqref{eqordtwojx}-\eqref{eqordtwojz} it can be seen that all components of the angular momentum are vanishing. Since all $\alpha_{cm}$ are zero, we still have freedom to make an Euler rotation to set $\alpha_{s1}=\alpha_{s-1}=\alpha_{s-2}=0$, which shows that we actually have a two-parameter family. Rather surprisingly, as we will show later, when we proceed to fifth order in the $\varepsilon$ expansion, the consistency conditions will restrict this prospective two-parameter family into two one-parameter families, an axisymmetric one with $\alpha_{s2}=0$ and a nonsymmetric one with $\alpha_{s0}=0$.

As the second case, we consider solutions with $\alpha_{c2}=0$ but $\alpha_{c0}\not=0$. In this case we can make an additional rotation to set, for example, $\alpha_{s1}=0$. Apart from special cases of the previously found two-parameter solution there is only one new solution in this case, which in its simplest form has only two nonzero amplitudes among $\alpha_{\sigma m}$, and these two have to agree, $\alpha_{c0}=\alpha_{s2}$, so it is a one-parameter family, which we call Solution C, since it will be considered in detail in Sec.~\ref{subsecoscill}. With a time shift of $-\pi/6$ and rotation $\pi/2$ around the $z$ axis we can transform it to a form where the two nonzero coefficients are $\alpha_{s0}=\alpha_{c2}$. To our knowledge, this solution has not been presented in the literature yet. From \eqref{eqordtwojz} and \eqref{eqordtwojy} it follows that the angular momentum is zero, so it is a nonrotating solution. We will see that it can be extended to higher order in the $\varepsilon$ expansion formalism. It would be important to show the existence of this family of solutions by direct numerical search, although this is not an easy task since it does not have any killing vector, so one has to consider the full $(3+1)$-dimensional problem numerically. Solution C, at least for small amplitudes, oscillates between an axially symmetric state with $\mathbb{S}_{20}$ angular dependence and a nonsymmetric state with $\mathbb{S}_{22}$ dependence, with a frequency close to $\omega=3$.

As the third case, we consider solutions with $\alpha_{c2}\not=0$ but $\alpha_{c0}=0$. In this class there are two new one-parameter families of solutions, and both are rotating solutions with one helical Killing vector. The first solution has only two equal nonzero components, $\alpha_{c2}=\alpha_{s-2}$. There is only one nonzero angular momentum, $J_z$, so the configuration is rotating around the $z$ axis. Since all metric components depend on $t$ and $\phi$ through trigonometric functions of integer multiples of $3t-2\phi$, the solution is rotating with angular frequency $\omega/m=3/2$ with respect to the coordinate $t$. This is the helically symmetric solution studied in detail in \cite{Dias12a,Horowitz14} and in \cite{Gregoire2017}, by both analytic and numerical methods. We will refer to it as Solution E in the following. More details about it will be given in Sec.~\ref{subsechelic22}.

The second new one-parameter family of solutions in the third case corresponds to a helically symmetric solution with $m=1$. However, since we have transformed away $\alpha_{c-1}$ and $\alpha_{c1}$, initially we find it in a nontrivial form. The solution has $\alpha_{s-2}=\alpha_{s0}=\alpha_{s2}=0$, and the remaining three constants are related by $\alpha_{c2}/\sqrt{2}=\alpha_{s-1}=\alpha_{s1}$. Calculating the angular momentum components, it turns out that $J_z=0$ and $J_x=J_y$ nonvanishing. Making an Euler rotation with $\alpha=7\pi/4$, $\beta=\pi/2$ and $\gamma=\pi$, we rotate the angular momentum into the direction of the $z$ axis, and we obtain the solution in its simplest form, where the only nonzero components are $\alpha_{c1}=\alpha_{s-1}$. After this rotation, only the $J_z$ angular momentum component is nonzero. This solution was included in the tables presented in \cite{Dias16a,Dias17a} as a likely geon solution, and its expansion up to fourth order was also performed in \cite{Dias17a}; however, no numerical study has been carried out to construct it. We refer to this family of solutions as Solution D in this paper. It will be discussed in Sec.~\ref{subsechelic21}. We will also show later that the consistency conditions can be satisfied at order $\varepsilon^5$, which strongly supports the claim that it is a true nonlinear one-parameter family. The $t$ and $\phi$ dependence of the metric is through trigonometric functions of multiples of $3t-\phi$, so the solution is rotating with angular frequency $\omega/m=3$.

In the fourth case, when both $\alpha_{c2}$ and $\alpha_{c0}$ are nonzero, somewhat surprisingly, there are no new solutions, only various rotated and time-shifted versions of those presented before. So performing the analysis to $\varepsilon^3$ order, we are left with three one-parameter families of solutions and with a fourth family that still has two independent parameters at this order.

\subsection{Prospective two-parameter family} \label{subsectwoparfam}

For the family of solutions that still appears to have two parameters at third order we choose a representation when only the two independent constants $\alpha_{c2}$ and $\alpha_{c0}$ can be nonzero from the ten parameters $\alpha_{\sigma m}$. We have proved earlier that all consistency conditions at third order in $\varepsilon$ can be satisfied in this case. However, at each resonant component, when a consistency condition may arise, we also have the freedom to add an arbitrary constant times the homogeneous solution to the generating function $\Phi$. At order $\varepsilon^3$, we denote by $c_{0}$ the new constant belonging to the scalar-type $l=2$, $m=0$ component, generated by the $\omega=3$ frequency source term. Similarly, there is another constant, denoted by $c_{2}$, from the $l=2$, $m=2$ component. There are other constants that are necessary to introduce at the other $\varepsilon^3$ components, but they will not modify the analysis that follows. There will be no consistency conditions to satisfy at fourth order in $\varepsilon$; however, one has to introduce a new constant $\nu_4$, giving the $\varepsilon^4$ order frequency change, in the $l=m=0$ component. At fifth order two consistency conditions will be important to us, the first from the $l=2$, $m=0$, $\omega=3$ scalar-type component and the second from the $l=2$, $m=2$, $\omega=3$ component,
\begin{align}
 \alpha_{c0}&\Big\{1836843008L^5\pi\left[1221(c_0\alpha_{c0}
  +c_2\alpha_{c2})-1792L^3\pi\nu_4\right] \notag \\
 &-6615\pi^2\left(1116206877\alpha_{c0}^4+2345476494\alpha_{c0}^2\alpha_{c2}^2
  +1078519297\alpha_{c2}^4\right) \\
 &+16\left(4588209459927\alpha_{c0}^4+9521613484590\alpha_{c0}^2\alpha_{c2}^2
  +4473144605015\alpha_{c2}^4\right) \Big\}=0 \ , \notag \\
 \alpha_{c2}&\Big\{1836843008L^5\pi\left[1221(c_0\alpha_{c0}
  +c_2\alpha_{c2})-1792L^3\pi\nu_4\right] \notag \\
 &-6615\pi^2\left(1172738247\alpha_{c0}^4+2157038594\alpha_{c0}^2\alpha_{c2}^2
  +1135050667\alpha_{c2}^4\right) \\
 &+16\left(4760806742295\alpha_{c0}^4+8946289210030\alpha_{c0}^2\alpha_{c2}^2
  +4645741887383\alpha_{c2}^4\right) \Big\}=0 \ . \notag
\end{align}
Multiplying the first equation by $\alpha_{c2}$, multiplying the second one by $\alpha_{c0}$, and subtracting, we obtain the following constraint for the two amplitudes:
\begin{equation}
 \alpha_{c0}\alpha_{c2}\left(\alpha_{c0}^2-3\alpha_{c2}^2\right)
 \left(3\alpha_{c0}^2-\alpha_{c2}^2\right)=0 \ . \label{eqc0c2cond}
\end{equation}
At least one of the four factors must vanish here.

If $\alpha_{c0}=0$, then we have a one-parameter family of nonrotating solutions that only has the $l=m=2$ component at the linear order. It is different from the helically symmetric rotating Solution E, which in addition to $l=m=2$ also has a time shifted $l=-m=2$ component. This new solution, which we call Solution B, will be discussed in Sec.~\ref{subsecnonrot22}. It has no continuous symmetries described by a Killing vector, so in order to construct it numerically one will need a $(3+1)$-dimensional code. To our knowledge, this solution has not been published in other papers yet. Since the spherical harmonic $\mathbb{S}_{22}$ can be rotated into $\mathbb{S}_{21}$ by an Euler rotation with $\alpha=\pi/4$, $\beta=\pi/2$ and $\gamma=0$, this solution can also be interpreted as the nonlinear generalization of the $l=2$, $m=1$ linear mode. It is also different from the helically rotating $m=1$ family (Solution D), because it does not have an $l=2$, $m=-1$ component.

If $\alpha_{c2}=0$, then we have an axisymmetric nonrotating solution, which has already been reported in the tables in \cite{Dias16a,Dias17a} as a likely geon solution. However, it was only calculated up to third order in $\varepsilon$, and no numerical study has been reported for its construction. We refer to this family as Solution A in the following. It will be discussed in Sec.~\ref{subsecaxial} in detail.

The third possibility in \eqref{eqc0c2cond} is $\alpha_{c0}^2=3\alpha_{c2}^2$. The case $\alpha_{c0}=-\sqrt{3}\alpha_{c2}$ can be transformed into the other case, satisfying $\alpha_{c0}=\sqrt{3}\alpha_{c2}$, with a rotation around the $z$ axis with angle $\alpha=\pi/2$. A linear solution satisfying $\alpha_{c0}=\sqrt{3}\alpha_{c2}\equiv\bar\alpha$ can be rotated by Euler angles $\alpha=\pi/2$, $\beta=\pi/2$ and $\gamma=\pi/2$ to a solution for which the only nonzero component is $\alpha_{c2}=2\bar\alpha$, so it is equivalent to the previously studied Solution B.

The fourth possibility in \eqref{eqc0c2cond} is $3\alpha_{c0}^2=\alpha_{c2}^2$. Here also, by a $\pi/2$ rotation around the axis $z$, we can transform the case $\sqrt{3}\alpha_{c0}=-\alpha_{c2}$ into $\sqrt{3}\alpha_{c0}=\alpha_{c2}$. Then a configuration with $\sqrt{3}\alpha_{c0}=\alpha_{c2}\equiv\bar\alpha$ can be rotated by Euler angles $\alpha=\pi$, $\beta=\pi/2$ and $\gamma=0$ to a solution that has the only component $\alpha_{c0}=-2\bar\alpha$, so it is equivalent to the axially symmetric Solution A. We note that the signature of any single $\alpha_{\sigma m}$ can be made opposite by a shift in the time coordinate $t$.

\section{Five families} \label{secfivefamily}

We have seen up to now that there are exactly five one-parameter families of solutions that have only $\omega=3$ frequency components in their small amplitude limit. In this section we present some more details about the expansion of these solutions up to fifth order in $\varepsilon$. Up to this order the function $\nu$ defined in \eqref{eqnudef} can be written as $\nu=1+\nu_2\varepsilon^2+\nu_4\varepsilon^4$. Using \eqref{eqomeganu}, the expansion of the physical frequency of the geon is
\begin{equation}
 \bar\omega=\frac{3}{L}\left(1+\omega_2\varepsilon^2+\omega_4\varepsilon^4\right)
 , \label{eqomegabarexp}
\end{equation}
where
\begin{equation}
 \omega_2=-\frac{1}{2}\nu_2 \quad , \qquad 
 \omega_4=\frac{3}{8}\nu_2^2-\frac{1}{2}\nu_4 \ . \label{eqom2om4}
\end{equation}
We write the expansion of the total mass and the angular momentum components as
\begin{align}
\frac{M}{L}&=M_2\varepsilon^2+M_4\varepsilon^4 \ , \label{eqmassexp} \\
\frac{J_i}{L^2}&=J_{i2}\varepsilon^2+J_{i4}\varepsilon^4 \ , \quad i=x,y,z\,.\label{eqjjiexp}
\end{align}
For each concrete configuration the coefficients of $M$ and $J_i$ can be calculated by the methods described in Appendix \ref{app:asymptotics}, using some algebraic manipulation software. Up to second order in $M$ Eq.~\eqref{eqmassexp} can be inverted as
\begin{equation}
 \varepsilon^2=\frac{1}{M_2}\frac{M}{L}
 -\frac{M_4}{M_2^3}\frac{M^2}{L^2} \ . \label{eqepsmpl}
\end{equation}

From the physical frequency $\bar\omega$ we can define a variable $\hat\varepsilon$ that is second order small in $\varepsilon$ by setting
\begin{equation}
 \bar\omega=\frac{3}{L}\left(1-\hat\varepsilon^2\right) \ , \label{eqbaromhateps}
\end{equation}
with inverse relation $\hat\varepsilon^2=1-L\bar\omega/3$. The expansion of this new variable is
\begin{equation}
 \hat\varepsilon^2=-\omega_2\varepsilon^2-\omega_4\varepsilon^4 \ ,
\end{equation}
where $\omega_2$ and $\omega_4$ are the same as in \eqref{eqomegabarexp}. This can be inverted to the given order to yield
\begin{equation}
 \varepsilon^2=-\frac{1}{\omega_2}\hat\varepsilon^2
 -\frac{\omega_4}{\omega_2^3}\hat\varepsilon^4 \ . \label{eqepsbyhateps}
\end{equation}
The variable $\hat\varepsilon$ is useful when giving physical characteristics of solutions, such as mass and angular momentum, since it is a quantity that is independent of the reparametrization freedom present in the variable $\varepsilon$. The parameter $\varepsilon$ has only a physical meaning yet to linear order. We can freely make the change $\varepsilon\to\tilde\varepsilon(\varepsilon)=\varepsilon+\sigma_3\varepsilon^3+\sigma_5\varepsilon^5+\ldots$, where $\sigma_3$ and $\sigma_5$ are arbitrary constants. 

We can expand the mass in terms of the reparametrization invariant $\hat\varepsilon$ parameter as
\begin{equation}
\frac{M}{L}=\hat M_2\hat\varepsilon^2+\hat M_4\hat\varepsilon^4 \ . \label{eqmasshatexp}
\end{equation}
Substituting \eqref{eqepsbyhateps} and \eqref{eqom2om4} into \eqref{eqmassexp} we can express the new type of coefficients in terms of the previous ones,
\begin{align}
 \hat M_2&=\frac{2}{\nu_2}M_2 \ , \label{eqhatmm2} \\
 \hat M_4&=\left(\frac{3}{\nu_2}-\frac{4\nu_4}{\nu_2^3}\right)M_2
 +\frac{4}{\nu_2^2}M_4 \ . \label{eqhatmm4}
\end{align}
Equation \eqref{eqmasshatexp} can be inverted as
\begin{equation}
 \hat\varepsilon^2=\frac{1}{\hat M_2}\frac{M}{L}
 -\frac{\hat M_4}{\hat M_2^3}\frac{M^2}{L^2} \ . \label{eqhatepsmpl}
\end{equation}

It is also useful to write the expansion of the physical frequency by using the mass as a small parameter,
\begin{equation}
 \bar\omega=\frac{3}{L}\left(1+\hat\omega_2\frac{M}{L}+\hat\omega_4\frac{M^2}{L^2}\right) .
 \label{eqbaromega}
\end{equation}
Substituting \eqref{eqhatepsmpl} into \eqref{eqbaromhateps}, we can see that the coefficients $\hat\omega_i$ can be expressed in terms of the $\hat M_i$ in \eqref{eqmasshatexp} in a simple way,
\begin{equation}
 \hat\omega_2=-\frac{1}{\hat M_2} \quad , \qquad
 \hat\omega_4=\frac{\hat M_4}{\hat M_2^3} \ . \label{eqomhatmhat}
\end{equation}

We can also expand the angular momentum components in terms of the mass,
\begin{equation}
 \frac{J_i}{L^2}=\hat J_{i2}\frac{M}{L}+\hat J_{i4}\frac{M^2}{L^2}
 +\hat J_{i6}\frac{M^3}{L^3} \ . \label{eqjimassexp}
\end{equation}
Substituting \eqref{eqepsmpl} into \eqref{eqjjiexp} we can get the first two coefficients in terms of the original coefficients in the $\varepsilon$ expansion,
\begin{align}
 \hat J_{i2}&=\frac{J_{i2}}{M_2} \ , \label{eqhatji2exp} \\
 \hat J_{i4}&=\frac{1}{M_2^2}\left(J_{i4}-\frac{J_{i2}}{M_2}M_4\right) \ . \label{eqhatji4exp}
\end{align}
To get the $J_{i6}$ coefficients directly, one should proceed to sixth order in the $\varepsilon$ expansion. However, there is an indirect method by using the first law of geon dynamics\cite{Dias12a},
\begin{equation}
 \frac{\mathrm{d}J}{\mathrm{d}M}=\frac{m}{\bar\omega} \ , \label{eqgeondyn}
\end{equation}
where $m$ is the spherical harmonic index of the linear seed, and $J$ is the total angular momentum. At present, this law is only a conjecture for the asymptotically AdS case, but the sketch of a proof is presented in \cite{Horowitz14}. We assume that the geon is rotated into a state satisfying $J_x=J_y=0$, such that $J=J_z$. Since $\bar\omega$ is known to second order in $M$ in \eqref{eqbaromega}, assuming the validity of relation \eqref{eqgeondyn} allows us to calculate one more term in \eqref{eqjimassexp}, without actually calculating the $\varepsilon$ expansion up to sixth order. Calculating the expansion of $1/\bar\omega$ from \eqref{eqbaromega}, and substituting into \eqref{eqgeondyn}, we obtain
\begin{align}
 \hat J_{z2}&=\frac{m}{3} \ , \label{eqhatjz2exp} \\
 \hat J_{z4}&=-\frac{m}{3}\hat\omega_2 \ , \label{eqhatjz4exp} \\
 \hat J_{z6}&=\frac{m}{9}\left(\hat\omega_2^2-\hat\omega_4\right) \ . \label{eqhatjz6exp}
\end{align}
For the two helically symmetric rotating geon configurations discussed in this paper we can easily check that the two methods give identical values for $\hat J_{z2}$ and $\hat J_{z4}$. For helically symmetric geons with $l=m=2$, i.e.\ Solution E in this paper, the validity of the conjectured first law of geon dynamics has been established to ${\cal O}(J^3)$ in Ref. \cite{Gregoire2017}. This makes us confident enough to use Eq.~\eqref{eqgeondyn} also for the other helically symmetric family, i.e.\ Solution D in the present paper.

\subsection{Axially symmetric \texorpdfstring{$(l,m)=(2,0)$}{(l,m)=(2,0)} solution} \label{subsecaxial}

We begin with the simplest family, the axisymmetric Solution A from Sec.~\ref{subsectwoparfam}, since this has the smallest number of spherical harmonic components in the expansion. Even at high orders it only contains $m=0$ components, no vector-type modes are generated, and there are no terms with $\sin(\omega t)$ time dependence. At linear order the solution is generated by the scalar mode $l=2$, $m=0$, which has the time dependence $\cos(3t)$. We take this mode with amplitude $\alpha^{Sc}_{200}\equiv\alpha_{c0}\equiv\alpha$. At second order in $\varepsilon$ there are $l=0,2,4$ source terms with static or $\cos(6t)$ time dependence. Since these terms are not resonant, the centrally and asymptotically well behaving generating function and metric can be constructed to $\varepsilon^2$ order. At the $l=m=0$ component a new parameter, $\nu_2$, is introduced, according to \eqref{nukdefeq}, \eqref{eqhttnubar} and \eqref{eqnubarnuk}, which determines the second order change in the oscillation frequency of the geon.

At third order there are $l=0,2,4,6$ source terms with $\cos(3t)$ and $\cos(9t)$ time dependence. The $\cos(9t)$ terms are resonant, but the consistency conditions are identically satisfied. Since the frequency $\omega=9$ is the triple of the basic frequency, the vanishing of these terms is analogous to the nonexistence of $(+++)$ terms in the expansion of spherically symmetric self-gravitating scalar fields with $\Lambda<0$ in \cite{Craps14}. It would be instructive to prove a similar statement for the nonspherical vacuum system studied in this paper. The $\cos(3t)$ term is resonant only for $l=2$, and gives the consistency condition
\begin{equation}
 \nu_2=\frac{1221}{3584}\,\frac{\alpha^2}{\pi L^4} \ . \label{eqnu2axysim}
\end{equation}
At each resonant source component with some $l$, $m$ and $\omega$, there is a non-negative integer $n$ such that $\omega=l+1+2n$, and we can introduce a new unspecified constant $c_{lm\omega}$ that corresponds to the amplitude $c_p$ of the homogeneous regular solution $p_1$ that we can add freely to the inhomogeneous solution according to \eqref{eqpintinhom}. For scalar-type perturbations $p_1$ is given by $p^{(S)}_{ln}$ in \eqref{pslnsol}, and for the vector type it is given by $p^{(V)}_{ln}$ in \eqref{eqpvlndef}. For the axially symmetric solution the introduction of the constants $c_{203}$, $c_{209}$, $c_{409}$ and $c_{609}$ are motivated by the appearance of source terms at third order. Rather surprisingly, we also have to introduce a new constant $c_{809}$, corresponding to the homogeneous solution $p^{(S)}_{80}$, even if there is no inhomogeneous term at $\varepsilon^3$ order with $l=8$. However, at order $\varepsilon^5$ there will be a consistency condition (coupled to other conditions) with $l=8$, $n=0$ and $\omega=9$, which can be solved only if $c_{809}$ is nonzero.

At fourth order in $\varepsilon$ there will be $l=0,2,4,6,8,10$ source terms with $\omega=0,6,12$ frequencies. Since these are nonresonant, the centrally regular asymptotically AdS perturbation can be calculated to this order. The constant $\nu_4$ describing the fourth order frequency change is introduced at the $l=m=0$ component. Having the metric up to fourth order in $\varepsilon$, we can calculate the first two nonzero coefficients in the expansion \eqref{eqmassexp} of the total mass $M$, yielding
\begin{align}
 M_2&=\frac{135}{512}\,\frac{\alpha^2}{L^4} \ , \label{eqaxysimm2}\\
 M_4&=\frac{135}{256}\,\frac{\alpha}{L^3}c_{203}
  +\frac{243(58800\pi^2-570653)}{102760448\pi}\,\frac{\alpha^4}{L^8} \ . \label{eqaxysimm4}
\end{align}
The expression for the coefficient $M_4$ involves the yet unknown constant $c_{203}$, but it is independent of the others.

At $\varepsilon^5$ order there are source terms with $l=0,2,4,6,8,10,12$ and $\omega=3,9,15$. The $\cos(15t)$ terms are resonant, but they do not give consistency conditions. The resonance conditions belonging to the $\cos(9t)$ terms at $l=2,4,6,8$ give four linear equations that can be solved for the unspecified constants that were introduced at third order:
\begin{align}
 c_{209}&={\frac{90140387102134562514971674773}{211426422177977512667270721536}}
  \,\frac{\alpha^3}{\pi L^5} \ , \\
 c_{409}&=-\frac{29367541304034979607272354161}{132141513861235945417044200960}
  \,\frac{\sqrt{5}\alpha^3}{\pi L^5} \ , \\
 c_{609}&=-\frac{34780824174291627596546643291}{10174896567315167797112403473920}
  \,\frac{\sqrt{5}\sqrt{13}\alpha^3}{\pi L^5} \ , \\
 c_{809}&=-\frac{100400488602602669643900951}{10333879326179467293942284778200}
  \,\frac{\sqrt{5}\sqrt{17}\alpha^3}{\pi L^5} \ .
\end{align}
We see here that it was really necessary to introduce the constant $c_{809}$. 

Because of the $\omega=l+1+2n$ condition, among the $\cos(3t)$ source terms only the one with $l=2$ is resonant at fifth order and gives the consistency condition
\begin{align}
 &1836843008\pi L^5\left(1792\pi L^3\nu_4-1221\alpha c_{203}\right) \notag\\
 &\ \ =27\left(2718938939216-273470684865\pi ^2\right)\alpha^4 \ , \label{eqaxyconcond}
\end{align}
which is the only restriction that involves $\nu_4$ and $c_{203}$. The reason that only a combination of these two constants gets determined is that we have not yet uniquely fixed how we parametrize the solutions in our one-parameter family by the variable $\varepsilon$. For a nonrotating family of solutions there are two natural ways to fix the parametrization, by connecting it either to the mass or to the oscillation frequency of the geon. The approach followed in \cite{Dias17a} to fix the reparametrization freedom is to set $M_4$ and all higher mass coefficients at zero, making the mass proportional to $\varepsilon^2$. An alternative way is to set $\omega_4$ and all higher coefficients zero in the expansion of $\bar\omega$ in \eqref{eqomegabarexp}.

The expansion parameters of the mass in terms of the $\hat\varepsilon$ parameter in \eqref{eqmasshatexp} can be calculated using \eqref{eqhatmm2}, \eqref{eqhatmm4}, and the consistency condition \eqref{eqaxyconcond},
\begin{align}
 \hat M_2&=\frac{630\pi}{407}\approx 4.8629 \ ,  \label{eqhatm2axys} \\
 \hat M_4&=\frac{27\pi(1476864425925\pi^2-14604436416496)}{617019996736}
 \approx -3.8999 \ . \label{eqhatm4axys}
\end{align}
Since both the mass and the parameter $\hat\varepsilon$ are independent of the parametrization choice for $\varepsilon$, these coefficients are independent of the constants $c_{203}$ and $\nu_4$. Although future higher order results and numerical analysis may change the picture, \eqref{eqhatm2axys} and \eqref{eqhatm4axys} indicate that the mass is likely to reach a maximum near the amplitude $\hat\varepsilon=0.79$, corresponding to frequency $\bar\omega=1.13/L$, and mass value $M=1.51L$. Higher amplitude states, where an increase in central amplitude corresponds to a total mass decrease, are expected to be unstable, which is a generic behavior for various astrophysical objects. This type of instability has been observed in \cite{Fodor15} for spherically symmetric self-gravitating scalar breathers with $\Lambda<0$.

Using \eqref{eqomhatmhat} we can calculate the expansion coefficients of the frequency $\bar\omega$ in terms of $M$ in \eqref{eqbaromega},
\begin{align}
 \hat\omega_2&=-\frac{407}{630\pi} \ , \\
 \hat\omega_4&=\frac{1476864425925\pi^2-14604436416496}{84756672000\pi^2} \ .
\end{align}
The value of $\hat\omega_2$ was already given in \cite{Dias17a}, but no fourth order results were presented there for this family of solutions. The numerical values of these constants are presented in Table \ref{tablehatomega}.
\begin{table}[!hbtp]
 \centering
 \begin{tabular}{|c|c|c|}
  \hline
  Solution & $\hat\omega_2$ & $\hat\omega_4$ \\
  \hline
  \hline
  A & -0.20564 & -0.033913 \\
  \hline
  B & -0.20564 & 0.034657 \\
  \hline
  C & -0.12311 & -0.12077 \\
  \hline
  D & -0.16112 & -0.0024852 \\
  \hline
  E & -0.27514 & 0.16970 \\
  \hline
 \end{tabular}
 \caption{Numerical values of $\hat\omega_2$ and $\hat\omega_4$ for the five families of solutions.}
 \label{tablehatomega}
\end{table}

\subsection{Nonrotating \texorpdfstring{$(l,m)=(2,2)$}{(l,m)=(2,2)} solution} \label{subsecnonrot22}

The second one-parameter family of solutions we consider is the one that reduces to a single $l=m=2$ mode to the linear order. This was named Solution B in Sec.~\ref{subsectwoparfam}. Since it does not include a time-shifted $l=-m=2$ mode, it is a nonrotating configuration. This can also be confirmed by calculating the angular momentum, which turns out to be zero. The solution does not have any continuous symmetries that correspond to a Killing vector field; in particular, it is not axially or helically symmetric. The existence of this type of AdS geon solution has not been reported in the literature before. We take the $l=m=2$ mode with amplitude $\alpha^{Sc}_{220}\equiv\alpha_{c2}\equiv\alpha$. At second order in $\varepsilon$ there are only scalar-type source terms with $(l,m)=(0,0)$, $(2,0)$, $(4,0)$ and $(4,4)$, with static or $\cos(6t)$ time dependence. There are no resonant terms and the centrally regular asymptotically AdS solution can be calculated.

At $\varepsilon^3$ order there are scalar-type source terms with $(l,m)=(2,2)$, $(4,2)$, $(6,2)$ and $(6,6)$, with $\cos(3t)$ and $\cos(9t)$ time dependence. There are also $(l,m)=(3,-2)$ vector-type source terms with $\sin(3t)$ and $\sin(9t)$ dependence. The $(l,m)=(2,2)$ resonance condition with $\omega=3$ yields the value for $\nu_2$ given in \eqref{eqnu2axysim}. All other resonance conditions are identically satisfied at this order. The resonant terms make necessary the introduction of the constants denoted by $c_{lm\omega}$ describing the amplitudes of the freely specifiable homogeneous solutions, of scalar-type $c_{223}$, $c_{229}$, $c_{429}$, $c_{629}$, $c_{669}$ and of vector-type $c_{3-29}$. However, it turns out that the fifth-order resonance conditions will be solvable only if one introduces five more constants at third order, which are not motivated by any resonant source terms, scalar-type $c_{829}$, $c_{869}$, and vector-type $c_{5-29}$, $c_{7-29}$, and $c_{7-69}$. It will turn out at order $\varepsilon^5$ that none of these $11$ constants are zero. According to our experience with several families of solutions, it appears to be a general principle that even if there are no source terms at $\varepsilon^3$ order for some values of $(l,m,\omega)$, it is necessary to allow a homogeneous solution with an unspecified amplitude for this mode at $\varepsilon^3$ order if there are resonance conditions at $\varepsilon^5$ order with these $(l,m,\omega)$ values.

At fourth order in $\varepsilon$ there are only nonresonant source terms with $l\leq 10$, $m\leq 8$, and with $\omega=0,6,12$. The equations for the $\varepsilon^4$ order metric perturbations can be solved. The expansion coefficient $M_2$ in \eqref{eqmassexp} is again given by \eqref{eqaxysimm2}, which is expected because of the identical factors in front of the parameters in \eqref{eqmasstenpar}. Rather surprisingly, the expansion coefficient $M_4$ also turns out to be exactly the same as that of the axisymmetric solution presented in \eqref{eqaxysimm4}. This is probably related to the fact that both of these one-parameter families of solutions originate from the conjectural two-parameter family discussed in Sec.~\ref{subsectwoparfam}. That family was valid only to third order in $\varepsilon$, and because of the consistency conditions at $\varepsilon^5$ order, only two one-parameter families survived from it.

At fifth order in $\varepsilon$ the $(l,m,\omega)=(2,2,3)$ resonance condition yields
\begin{align}
 &1836843008\pi L^5\left(1792\pi L^3\nu_4-1221\alpha c_{223}\right) \notag\\
 &\ \ =\left(74331870198128-7508360162205\pi ^2\right)\alpha^4 \ . \label{eql2m2nrcond}
\end{align}
There are ten more consistency conditions belonging to $\omega=9$ resonance terms, which can be solved uniquely for the ten remaining $c_{lm\omega}$ constants. Because of the large number of digits appearing in the rational numbers we do not present their detailed values here.

The expansion parameter $\hat M_2$ of the mass in terms of the small parameter $\hat\varepsilon$ in \eqref{eqmasshatexp} is the same as in \eqref{eqhatm2axys}, but $\hat M_4$ turns out to be different, because of the difference between the right-hand sides of the consistency conditions \eqref{eqaxyconcond} and \eqref{eql2m2nrcond},
\begin{equation}
 \hat M_4=\frac{\pi(40498597854225\pi^2-398922377441872)}{617019996736}
 \approx 3.985 \ . \label{eqhatm4l2m2nr}
\end{equation}
Surprisingly, even the signature of $\hat M_4$ is the opposite, so in this case we have no indication for a maximum in the mass. This picture may change at higher orders in the expansion, and numerical analysis would be necessary to decide whether a maximal mass exists for this family of solutions. The coefficients $\hat\omega_i$ in the expansion \eqref{eqbaromega} of the frequency in terms of the mass can be calculated by the general formulas \eqref{eqomhatmhat}, and their numerical values are given in Table \ref{tablehatomega}.

\subsection{Solution oscillating between \texorpdfstring{$(l,m)=(2,2)$}{(l,m)=(2,2)} and \texorpdfstring{$(2,0)$}{(2,0)} states} \label{subsecoscill}

We continue with Solution C from Sec.~\ref{subsecthirdord}. This one-parameter family of solutions has two components in the linear order in $\varepsilon$, one $(l,m)=(2,2)$ component with $\cos(3t)$ time dependence, and another $(2,0)$ component with $\sin(3t)$ dependence. The two modes must have equal amplitudes but opposite time phases, $\alpha^{Sc}_{220}\equiv\alpha_{c2}\equiv\alpha$ and $\alpha^{Ss}_{200}\equiv\alpha_{s0}\equiv\alpha$. Similar to the previous family, this is also a nonrotating solution with zero angular momentum and no continuous symmetries, and it has not been reported in the literature before. This is the technically most complicated family among the five solutions considered in this paper, when considering the number of components and the length of the expressions.

At second order in $\varepsilon$ there are $(l,m)=(0,0)$, $(2,0)$, $(4,0)$, $(4,2)$ and $(4,4)$ scalar-type source terms with $\omega=0$ or $6$ frequency, and one static vector-type term with $l=3$, $m=-2$. At $\varepsilon^3$ order there are scalar-type source terms belonging to $l=0,2,4,6$ and $m=0,2,4,6$ with $\omega=3,9$. There are also vector-type source terms with $l=3,5$ and $m=-2,-4$. The $(l,m,\omega)=(2,0,3)$ and $(2,2,3)$ components give the same consistency condition,
\begin{equation}
 \nu_2=\frac{731}{1792}\,\frac{\alpha^2}{\pi L^4} \ .
\end{equation}
All other consistency conditions are trivially satisfied. Because of the presence of homogeneous solutions, the following nine $c_{lm\omega}$ constants have to be introduced: $c_{203}$, $c_{223}$, $c_{409}$, $c_{449}$, $c_{609}$, $c_{629}$, $c_{649}$, $c_{669}$ and $c_{3-29}$. To be able to solve the fifth order conditions one also has to introduce five more constants: $c_{809}$, $c_{849}$, $c_{889}$, $c_{7-29}$ and $c_{7-69}$. All of these will take a nonzero value from the fifth order consistency conditions.

The equations for the $\varepsilon^4$ order metric perturbations can be solved. The expansion coefficients of the total mass $M$ in terms of $\varepsilon$ in \eqref{eqmassexp} turn out to be
\begin{align}
 M_2&=\frac{135}{256}\,\frac{\alpha^2}{L^4} \ , \label{eqm20m2}\\
 M_4&=\frac{135}{256}\,\frac{\alpha}{L^3}(c_{203}+c_{223})
  +\frac{729(39200\pi^2-385767)}{51380224\pi}\,\frac{\alpha^4}{L^8} \ . \label{eqm20m4}
\end{align}
The value of $M_2$ here is the double of that in \eqref{eqaxysimm2}, since we have now included two modes with amplitude $\alpha$ at first order in $\varepsilon$.

At fifth order in $\varepsilon$ there will be two conditions coming from the $(l,m,\omega)=(2,0,3)$ and $(2,2,3)$ source terms. These are equivalent to
\begin{align}
 &c_{203}=c_{223} \ , \\
 &7347372032\pi L^5\left(896\pi L^3\nu_4-731\alpha c_{203}\right) \notag\\
 &\ \ =\left(491118296077952-49227041244465\pi ^2\right)\alpha^4 \ .
\end{align}
There will be $20$ more nontrivial consistency conditions coming from various $\omega=9$ source terms. It turns out that these constraints are not linearly independent, since they can be uniquely solved for the remaining twelve $c_{lm9}$ constants.

The expansion coefficients of the mass in terms of the parameter $\hat\varepsilon$ in \eqref{eqmasshatexp} can be calculated using \eqref{eqhatmm2} and \eqref{eqhatmm4},
\begin{align}
 \hat M_2&=\frac{1890\pi}{731}\approx 8.1226 \ ,  \label{eqhatm2m20} \\
 \hat M_4&=\frac{27\pi(260296835303925\pi^2-2590848003556096)}{28599479507456}
 \approx -64.719 \ . \label{eqhatm4m20}
\end{align}
This indicates that the mass is likely to reach a maximum near the amplitude $\hat\varepsilon=0.25$, corresponding to frequency $\bar\omega=2.8/L$, and mass value $M=0.25L$. The coefficients $\hat\omega_i$ in expansion \eqref{eqbaromega} of the frequency in terms of $M/L$ can be calculated by general formulas \eqref{eqomhatmhat}, and they are given in Table \ref{tablehatomega}.

\subsection{Helically symmetric \texorpdfstring{$(l,m)=(2,\pm 1)$}{(l,m)=(2,+-1)} solution} \label{subsechelic21}

The next solution we consider in detail is Solution D from Sec.~\ref{subsecthirdord}. It is a one-parameter family of helically symmetric configurations, which in the small amplitude limit reduces to the linear combination of two components, an $l=2$, $m=1$ mode with $\cos(3t)$ time dependence, and an $l=2$, $m=-1$ mode with $\sin(3t)$ time dependence. The nonzero amplitudes are $\alpha^{Sc}_{210}\equiv\alpha_{c1}\equiv\alpha$ and $\alpha^{Ss}_{2-10}\equiv\alpha_{s-1}\equiv\alpha$. Since the two modes must have identical amplitudes, the $\varepsilon$ order metric perturbation components are proportional to $\cos(3t-\phi)$ or $\sin(3t-\phi)$. Even at higher orders, the $t$ and $\phi$ dependence will only be through trigonometric functions of integer multiples of $3t-\phi$, so the solution is rotating with angular velocity $3$ with respect to the time coordinate $t$. This solution was already reported in \cite{Dias16a,Dias17a} as a possible geon configuration, where a third order analysis in $\varepsilon$ was performed for it. Here we show that the fifth order consistency conditions can also be satisfied, which makes it extremely likely that it corresponds to an actual solution of the nonlinear equations, and we present higher order results for its frequency, mass, and angular momentum.

At each order in $\varepsilon$, and for arbitrary $l$, if we have a mode with some given $m$ value, then because of the helical symmetry, this mode can only have a special time dependence. If $m\geq 0$ then the time dependence must be $\cos(3mt)$, and if $m<0$, then $\sin(-3mt)$. In particular, $m=0$ modes must be static. Because of this, we do not explicitly state the frequency of the various modes in the following. Furthermore, the amplitudes of the $m$ and $-m$ modes will always be identical.

If at some resonant source term corresponding to $(l,m)$ we add the homogeneous solution with amplitude $c_{lm}$, then the helical symmetry is preserved only if at the $(l,-m)$ component we add the corresponding homogeneous solution with the same amplitude. If we would introduce independent $c_{l-m}$ constants for the $-m$ mode, then the consistency conditions at fifth order would imply that $c_{l-m}=c_{lm}$, ensuring the helical symmetry of the full nonlinear solution. The resonant source terms at $\varepsilon^3$ order motivate the introduction of the following amplitude constants: $c_{21}$, $c_{43}$, $c_{63}$ coming from scalar-type terms, and $c_{33}$ coming from a vector-type term. The resonance conditions at $\varepsilon^5$ order can be solved only if one introduces three more constants giving the amplitudes of homogeneous solutions not motivated by any source terms, scalar-type $c_{83}$ and vector-type $c_{53}$, $c_{73}$.

At $\varepsilon^4$ order there are scalar-type source terms with $l=0,2,4,6,8,10$ and $m=0,\pm 2,\pm 4$, as well as vector-type ones with $l=1,3,5,7,9$ and $m=0,\pm 2,\pm 4$. There are no conditions, and the equations can be solved for the metric perturbation. The value of the coefficient $M_2$ in the expansion \eqref{eqmassexp} of the mass is the same as in \eqref{eqm20m2}. The next coefficient is
\begin{equation}
 M_4=\frac{135}{128}\,\frac{\alpha}{L^3}c_{21}
  +\frac{135(8400\pi^2-83201)}{1835008\pi}\,\frac{\alpha^4}{L^8} \ .
\end{equation}
Only the angular momentum component $J_z$ is nonzero, and its expansion coefficients in \eqref{eqjjiexp} are
\begin{align}
 J_{z2}&=\frac{45}{256}\,\frac{\alpha^2}{L^4} \ , \label{eqhel21jz2} \\
 J_{z4}&=\frac{45}{128}\,\frac{\alpha}{L^3}c_{21}
  +\frac{15(25200\pi^2-246733)}{1835008\pi}\,\frac{\alpha^4}{L^8} \ .
\end{align}

At $\varepsilon^5$ order there are scalar terms with $l=2,4,6,8,10,12$ and $m=\pm 1,\pm 3,\pm 5,\pm 7$, and there are vector-type ones with $l=1,3,5,7,9,11$ and $m=\pm 1,\pm 3,\pm 5,\pm 7$. The $(l,m)=(2,1)$ condition takes the form
\begin{align}
 &57401344\pi L^5\left(192\pi L^3\nu_4-205\alpha c_{21}\right) \notag\\
 &\ \ =3\left(361971753894-36598980775\pi ^2\right)\alpha^4 \ .
\end{align}
The $m=3$ scalar and vector resonance conditions for $l=3,4,5,6,7,8$ uniquely determine the six $c_{l3}$ constants, and there are no more nontrivial restrictions at fifth order.

According to \eqref{eqmasshatexp}, \eqref{eqhatmm2} and \eqref{eqhatmm4}, the expansion coefficients of the mass in terms of the parameter $\hat\varepsilon$ are
\begin{align}
 \hat M_2&=\frac{81\pi}{41}\approx 6.2066 \ ,   \\
 \hat M_4&=\frac{729\pi(38897276775\pi^2-383950851034)}{193171778800}
 \approx -0.59418 \ .
\end{align}
This would indicate that the mass is likely to reach a maximum near the amplitude $\hat\varepsilon=2.3$ with mass value $M=16.2L$. This high value of $\hat\varepsilon=2.3$ would correspond to a negative $\bar\omega$ frequency, so we need higher order analysis and numerical calculations to provide information about the existence and possible value of the mass maximum. The coefficients $\hat\omega_i$ in the expansion \eqref{eqbaromega} of the frequency in terms of $M/L$ can be calculated by the general formulas \eqref{eqomhatmhat}, and their rounded values are included in Table \ref{tablehatomega}.

Only the $z$ component of the angular momentum is nonzero, and its expansion coefficients in terms of the mass in \eqref{eqjimassexp} are
\begin{align}
 \hat J_{z2}&=\frac{1}{3} \ , \\
 \hat J_{z4}&=\frac{41}{486\pi} \ , \\
 \hat J_{z6}&=\frac{3460269166106-350075490975\pi^2}{165502537200\pi^2} \ .
\end{align}
The coefficients $\hat J_{z2}$ and $\hat J_{z4}$ can be obtained by direct calculation from the metric either using \eqref{eqhatji2exp} and \eqref{eqhatji4exp} or using the conjectured first law of geon dynamics by \eqref{eqhatjz2exp} and \eqref{eqhatjz4exp}, leading to the same results. Since we have not calculated the metric up to sixth order for this geon family, the value of $\hat J_{z6}$ was only calculated using the first law by \eqref{eqhatjz6exp}.

\subsection{Helically symmetric \texorpdfstring{$(l,m)=(2,\pm 2)$}{(l,m)=(2,+-2)} solution} \label{subsechelic22}

The last solution among the five families that we consider in detail is Solution E from Sec.~\ref{subsecthirdord}. Similar to Solution D studied in the previous subsection, this is a one-parameter family of helically symmetric configurations, which in the small amplitude limit reduces to the linear combination of two components, now an $l=2$, $m=2$ mode with $\cos(3t)$ time dependence, and an $l=2$, $m=-2$ mode with $\sin(3t)$ time dependence. The two amplitudes must be identical, $\alpha^{Sc}_{220}\equiv\alpha_{c2}\equiv\alpha$ and $\alpha^{Ss}_{2-20}\equiv\alpha_{s-2}\equiv\alpha$. The $\varepsilon$ order metric perturbation components are proportional to $\cos(3t-2\phi)$ or $\sin(3t-2\phi)$. At higher orders the $t$ and $\phi$ dependence will be only through trigonometric functions of integer multiples of $3t-2\phi$, so the solution is rotating with angular velocity $3/2$ with respect to the time coordinate $t$. This is the most thoroughly studied AdS geon solution in the literature, and it has been investigated in detail in \cite{Dias12a,Horowitz14} and in \cite{Gregoire2017}, by both analytic and numerical methods.

At each order, and for arbitrary $l$, if $m\geq 0$, then the time dependence of the mode must be $\cos(3mt/2)$, and if $m<0$ then $\sin(-3mt/2)$. In particular, $m=0$ modes must be static, and there are no modes with odd $m$ in this case. The amplitudes of the $m$ and $-m$ modes will always be identical.

At second order in $\varepsilon$ there are scalar-type source terms with $(l,m)=(0,0)$, $(2,0)$, $(4,0)$, $(4,\pm 4)$, and vector-type terms with $(1,0)$, $(3,0)$. The unique centrally regular and asymptotically AdS metric perturbation can be calculated.

At $\varepsilon^3$ order there are scalar-type source terms with $(l,m)=(2,\pm 2)$, $(4,\pm 2)$, $(6,\pm 2)$, $(6,\pm 6)$, and vector-type terms with $(3,\pm 2)$, $(5,\pm 2)$. The $(l,m)=(2,\pm 2)$ components yield the resonance condition
\begin{equation}
 \nu_2=\frac{4901}{5376}\,\frac{\alpha^2}{\pi L^4} \ .
\end{equation}
All other conditions are trivially satisfied at this order. The presence of source terms motivate the introduction of the two scalar-type $c_{lm}$ constants, $c_{22}$ and $c_{66}$, corresponding to the unspecified amplitudes of homogeneous solutions. The fifth order consistency conditions can be solved only if a scalar-type $c_{86}$ and a vector-type $c_{76}$ constant is also introduced. Although this is the AdS geon solution that has been studied to the most detail in the literature, the necessity of these additional homogeneous solutions with nonzero amplitudes has not been reported up to now.

At $\varepsilon^4$ order there are scalar-type source terms with $l=0,2,4,6,8,10$ and $m=0,\pm 4,\pm 8$, and vector-type ones with $l=1,3,5,7,9$ and $m=0,\pm 4,\pm 8$. There are no conditions, and the equations can be solved for the metric perturbation. The value of the coefficient $M_2$ in the expansion \eqref{eqmassexp} of the mass is again the same as in \eqref{eqm20m2}. The next coefficient is
\begin{equation}
 M_4=\frac{135}{128}\,\frac{\alpha}{L^3}c_{22}
  +\frac{27(1528800\pi^2-15345433)}{51380224\pi}\,\frac{\alpha^4}{L^8} \ .
\end{equation}
Only the angular momentum component $J_z$ is nonzero, and its expansion coefficients in \eqref{eqjjiexp} are
\begin{align}
 J_{z2}&=\frac{45}{128}\,\frac{\alpha^2}{L^4} \ , \\
 J_{z4}&=\frac{45}{64}\,\frac{\alpha}{L^3}c_{22}
  +\frac{3(4586400\pi^2-45350159)}{25690112\pi}\,\frac{\alpha^4}{L^8} \ .
\end{align}
The coefficient $J_{z2}$ is double that of the value for the $m=1$ helical family, presented in \eqref{eqhel21jz2}.

At $\varepsilon^5$ order there are scalar source terms with $l=0,2,4,6,8,10,12$ and $m=\pm 2,\pm 6,\pm 10$, and there are vector-type ones with $l=3,5,7,9,11$ and $m=\pm 2,\pm 6,\pm 10$. The $(l,m)=(2,2)$ condition takes the form
\begin{align}
 &7347372032\pi L^5\left(2688\pi L^3\nu_4-4901\alpha c_{22}\right) \notag\\
 &\ \ =15\left(324432451551360-32944117238417\pi ^2\right)\alpha^4 \ . \label{eqfifthordc22}
\end{align}
The $(l,m)=(6,6)$, $(8,6)$ scalar and $(l,m)=(7,6)$ vector resonance conditions determine the yet unspecified constants
\begin{align}
 c_{66}&={\frac{342951235065187003920120571}{1274223922034023348070582099200}}
  \,\frac{\sqrt{2}\sqrt{5}\sqrt{7}\sqrt{11}\sqrt{13}\alpha^3}{\pi L^5} \ , \\
 c_{76}&={\frac{8354315114469291547881297}{22753998607750416929831823200}}
  \,\frac{\sqrt{3}\sqrt{11}\sqrt{13}\alpha^3}{\pi L^5} \ , \\
 c_{86}&=-{\frac{3850991730510763011266813}{11604539289952712634214229832000}}
  \,\frac{\sqrt{2}\sqrt{5}\sqrt{11}\sqrt{13}\sqrt{17}\alpha^3}{\pi L^5} \ .
\end{align}

The expansion coefficients of the mass in \eqref{eqmasshatexp} in terms of the parameter $\hat\varepsilon$ can be calculated using \eqref{eqhatmm2}, \eqref{eqhatmm4}, and the fifth order constraint \eqref{eqfifthordc22},
\begin{align}
 \hat M_2&=\frac{5670\pi}{4901}\approx 3.6345 \ ,   \\
 \hat M_4&=\frac{729\pi(869318078838825\pi^2-8549162771834624)}{8619064008828416}
 \approx 8.1476 \ .
\end{align}
Since $M_4>0$ we cannot infer a mass maximum from these values. The coefficients $\hat\omega_i$ in the expansion \eqref{eqbaromega} of the frequency in terms of $M/L$ can be calculated by \eqref{eqomhatmhat}. The numerical values for $\hat\omega_2$ and $\hat\omega_4$ for the five families discussed in this paper are given in Table \ref{tablehatomega}.

For this solution, only the $z$ component of the angular momentum is nonzero. The expansion coefficients of $J_z$ in terms of the mass in \eqref{eqjimassexp} are
\begin{align}
 \hat J_{z2}&=\frac{2}{3} \ , \label{eqjz2hel22} \\
 \hat J_{z4}&=\frac{4901}{17010\pi} \ , \label{eqjz4hel22} \\
 \hat J_{z6}&=\frac{77065569309012736-7823862709549425\pi^2}{741451366656000\pi^2}
  \ . \label{eqjz6hel22}
\end{align}
The direct calculation by \eqref{eqhatji2exp} and \eqref{eqhatji4exp} gives the same result for $\hat J_{z2}$ and $\hat J_{z4}$ as the first law of geon dynamics using \eqref{eqhatjz2exp} and \eqref{eqhatjz4exp}. Since the $\varepsilon$ expansion of this solution was presented in this subsection only up to fifth order, the above results allow the direct expansion of $J_z$ only up to ${\cal O}(M^2)$. The calculation of $\hat J_{z6}$ from this information is only possible by the first law method using \eqref{eqhatjz6exp}. In Ref. \cite{Gregoire2017}, however, a sixth order computation in $\epsilon$ has already been carried out to obtain the mass-angular momentum relation to ${\cal O}(J^3)$; cf.\ Eq.\ (20c) of that paper. That equation is precisely the inverse of the angular momentum-mass relation \eqref{eqjimassexp}, up to third order in $J$ or $M$, with the coefficients given in \eqref{eqjz2hel22}-\eqref{eqjz6hel22}. The direct computation of Eq.\ (20c) of Ref. \cite{Gregoire2017} made it possible to verify the validity of the conjectured first law of geon dynamics, Eq.\eqref{eqgeondyn}, up to ${\cal O}(J^3)$.

\section{Conclusions}

In the present paper we have given a detailed construction of five one-parameter families of asymptotically AdS time-periodic vacuum geon solutions with $\omega=3$, by performing a fifth order perturbative analysis. We have made an effort to provide all necessary technical details in order to make the checking of the results relatively easy, as well as to make our techniques available for other researchers. AdS geon solutions have already been studied in the literature, but many technical details have not been published yet. Furthermore, up to now, only helically rotating or axially symmetric geon solutions have been found. By our high order perturbative method we have shown the existence of two different one-parameter families of solutions that do not have any continuous symmetries. Surprisingly, they are not rotating, since all their angular momentum components are zero. For the first time, we have considered in detail the consistency conditions that appear at fifth order in the expansion. We have shown that they can be solved only if some nontrivial homogeneous solutions are added at third order, which are not motivated by the presence of any inhomogeneous terms at third order.

The most important result of our work is that the presented five families are exactly the ones that reduce to only $\omega=3$ frequency modes at linear order, and there are no other such solutions. We have considered $\omega=3$, because it is the lowest possible frequency in this system. One may be tempted to easily state that the number of these families obviously agrees with the multiplicity of the $l=2$ modes, spanning $-2\leq m\leq 2$. However, each of these modes may have $\cos$ or $\sin$ time dependence, which seemingly doubles the number of possible modes. On the other hand, spatial rotation and time shift can make apparently different solutions identical. For example the $l=2$, $m=2$ spherical harmonic mode can easily be rotated to an $l=2$, $m=1$ state.

Our high order perturbative results not only establish the existence of these families, but also allow us to calculate how their frequency, mass, and angular momentum depend on each other. We have obtained expressions for these dependencies that are valid to higher orders than earlier results available in the literature. These relations will likely be useful when later direct numerical searches will be performed for these configurations, making it possible to check the consistency of the two kinds of results. The next big challenge will be the investigation of the stability of these geon solutions by a numerical time-evolution code. Numerical studies of nonspherically symmetric evolution of scalar fields with AdS asymptotics has been reported recently in Refs.~\cite{Bantilan17,Choptuik17}. An important open question about AdS geons is whether all of the one-parameter families presented in this paper are stable up to a state where they reach a mass maximum.

\section*{Acknowledgments}

We are very grateful to Philippe Grandcl\'ement and Gr\'egoire Martinon for their kind help and frequent discussions when we were visiting Paris Observatory in Meudon. We also acknowledge useful discussions with Andrzej Rostworowski. G.~F.~is grateful for the kind hospitality of the LUTH research group at the Paris Observatory in Meudon during his two-year Marie Curie fellowship. This research has been supported in part by OTKA Grant No.~K 101709 and by the Marie Curie Actions Intra European Fellowship of the European Community’s Seventh Framework Programme under Contract No.~PIEF-GA-2013-621992.

\appendix

\section{Asymptotics and conserved quantities} \label{app:asymptotics}

For the treatment of asymptotically anti-de Sitter spacetimes we use the definition based on Penrose's conformal treatment of infinity, which was first proposed for the $\Lambda<0$ case in \cite{Hawking1983175} and was investigated in more detail in \cite{AshMag,AshDas,Hollands2005}. Our manifold $(\mathcal{M},g_{\mu\nu})$ can be asymptotically AdS if there exist a manifold $(\mathcal{\tilde M},\tilde g_{\mu\nu})$ with boundary $\mathscr{I}$, and a diffeomorphism from $\mathcal{M}$ onto $\mathcal{\tilde M}\setminus\mathscr{I}$, such that $\tilde g_{\mu\nu}=\Omega^2g_{\mu\nu}$. It is also required that $\mathscr{I}$ is topologically $\mathcal{S}^2\times\mathbb{R}$, and that $\Omega=0$ on $\mathscr{I}$, but its gradient is nonvanishing there. Using the radial coordinate $x$, it is easy to see that our one-parameter family of solutions represented by \eqref{gmunuexp1} with the boundary conditions \eqref{nukdefeq} and \eqref{eqgmunulim} clearly satisfies these requirements. The manifold $\mathcal{M}$ belongs to the region $0\leq x<\pi/2$, and $\mathscr{I}$ is the surface $x=\pi/2$. Choosing the conformal factor as in \eqref{eqconffact}, the conformally transformed metric $\tilde g_{\mu\nu}=\Omega^2 g_{\mu\nu}$ is regular at $\mathscr{I}$. The spacelike one-form $n_\mu=\tilde\nabla_\mu\Omega$ is orthogonal to $\mathscr{I}$, and has the norm $\sqrt{\tilde g^{\mu\nu}n_\mu n_\nu}=1/L$ on $\mathscr{I}$.

In the case of four spacetime dimensions there is an additional condition that asymptotically AdS spacetimes have to satisfy. This condition has various equivalent forms. The first form of the condition is to require that the induced metric, $\tilde\gamma_{\mu\nu}=\tilde g_{\mu\nu}-L^2 n_\mu n_\nu$, on the hypersurface $\mathscr{I}$ is conformally flat. Since we have required the boundary conditions \eqref{nukdefeq} and \eqref{eqgmunulim}, the induced metric on the timelike surface corresponding to infinity in the $(t,\theta,\phi)$ coordinate system is
\begin{equation}
 \tilde\gamma_{\mu\nu}=\text{diag}\left(
 -\nu  \ , \ 1  \ , \ \sin^2\theta \right) \ ,
\end{equation}
where $\nu$ is the $\varepsilon$ dependent constant defined in \eqref{eqnudef}. Apart from the rescaling of the time coordinate, this metric is the same as for the exact AdS case, ensuring the asymptotically AdS property of the investigated family of solutions.

A second equivalent way to state the conformal flatness condition is to require that the conformal group of $\mathscr{I}$, with respect to the metric induced on it, is the anti-de Sitter group. A third equivalent possibility, in terms of four-dimensional language, is to require the vanishing of the magnetic part of the asymptotic Weyl curvature on $\mathscr{I}$.

Similar to the calculation of the curvature belonging to the physical metric $g_{\mu\nu}$ in the form \eqref{eqgmunufinite} in Sec.~\ref{sec-highord}, we can calculate the Weyl tensor $\tilde C_{\mu\nu\rho\sigma}$ belonging to the conformal metric $\tilde g_{\mu\nu}$ by some algebraic manipulation software up to a given order in $\varepsilon$. The dual Weyl tensor ${}^*\tilde C_{\mu\nu\rho\sigma}=\tilde\epsilon_{\mu\nu\alpha\beta}\tilde C^{\alpha\beta}_{\ \ \rho\sigma}$ can also be calculated, where $\tilde\epsilon_{\mu\nu\alpha\beta}$ is the totally antisymmetric tensor belonging to $\tilde g_{\mu\nu}$. In four spacetime dimensions, from the asymptotically AdS conditions it follows that the Weyl tensor (and its dual) is vanishing on $\mathscr{I}$. Consequently, one can define the leading order asymptotic Weyl and dual Weyl tensors on $\mathscr{I}$ by
\begin{equation}
 K_{\mu\nu\rho\sigma}=\lim_{\to\mathscr{I}}\frac{1}{\Omega}\,\tilde C_{\mu\nu\rho\sigma}
 \quad , \qquad
 {}^*K_{\mu\nu\rho\sigma}=\lim_{\to\mathscr{I}}\frac{1}{\Omega}
 \,{}^*\tilde C_{\mu\nu\rho\sigma} \ .
\end{equation}
The electric and the magnetic parts of the asymptotic Weyl curvature on $\mathscr{I}$ are defined as
\begin{equation}
 \mathcal{E}_{\mu\nu}=L^2 K_{\mu\rho\nu\sigma}n^\rho n^\sigma \quad , \qquad
 \mathcal{B}_{\mu\nu}=L^2\, {}^*K_{\mu\rho\nu\sigma}n^\rho n^\sigma \ .
\end{equation}
Because of the antisymmetry of the Weyl tensor, $\mathcal{E}_{\mu\nu}n^\nu=\mathcal{B}_{\mu\nu}n^\nu=0$, so these are symmetric tensors in the timelike hypersurface $\mathscr{I}$.

The vanishing of the magnetic part is equivalent to the conformal flatness condition in the definition of asymptotically AdS spacetimes, assuming that Einstein's equations hold\cite{AshMag,AshDas}. Hence from our boundary conditions \eqref{nukdefeq} and \eqref{eqgmunulim} it necessarily follows that $\mathcal{B}_{\mu\nu}=0$. We have also checked this by direct calculation up to certain order in $\varepsilon$ for several concrete geon configurations.

Since $\tilde\gamma^{\mu\nu}\mathcal{E}_{\mu\nu}=0$, and without matter fields $\tilde D^\mu\mathcal{E}_{\mu\nu}=0$, where $\tilde D_\mu$ is the derivative operator belonging to $\tilde\gamma^{\mu\nu}$, the electric part can be used to define conserved quantities. Asymptotic symmetries on the spacetime correspond to conformal Killing fields $\xi^\mu$ on $\mathscr{I}$. For every choice of $\xi^\mu$ and a 2-sphere cross section $\mathcal{C}$ of $\mathscr{I}$ a conserved quantity is defined in \cite{AshMag},
\begin{equation}
 Q_\xi=-\frac{L}{8\pi}\oint_\mathcal{C}\mathcal{E}_{\mu\nu}\xi^\mu u^\nu\mathrm{d}\tilde S \ ,
\end{equation}
where $\mathrm{d}\tilde S$ is the volume element on $\mathcal{C}$, and $u^\nu$ is the unit normal to $\mathcal{C}$, both with respect to the metric $\tilde\gamma_{\mu\nu}$. When there are no matter fields, $Q_\xi$ is independent of the choice of the hypersurface $\mathcal{C}$, so it is absolutely conserved. The simplest choice for $\mathcal{C}$ is to take a constant $t$ section, and then $u^\mu=(1/\sqrt{\nu},0,0)$. Then the conserved charge can be calculated as
\begin{equation}
 Q_\xi=-\frac{L}{8\pi}\int_0^{2\pi}\mathrm{d}\phi\int_0^\pi\mathrm{d}\theta\sin\theta\frac{1}{\sqrt{\nu}}\mathcal{E}_{\mu t}\xi^\mu \ .
\end{equation}

To calculate the total mass $M\equiv Q_\xi$ of the configuration we need to use the $\xi^\mu$ conformal Killing vector on $\mathscr{I}$ corresponding to the timelike asymptotic Killing vector that at large distances tends to the timelike Killing vector $\mathrm{d}/\mathrm{d}\bar t$. This means that we have to use the conformal Killing vector $\xi^\mu=(1/(L\sqrt{\nu}),0,0)$ for the mass calculation, where the coordinates are $(t,\theta,\phi)$ now. To calculate the $J_x$, $J_y$, and $J_z$ components of the angular momentum we choose the conformal Killing vectors as $\xi^\mu=(0,\sin\phi,\cot\theta\cos\phi)$, $\xi^\mu=(0,-\cos\phi,\cot\theta\sin\phi)$, and $\xi^\mu=(0,0,-1)$, respectively. For the three Killing vectors inducing the angular momentum components we have included an extra $-1$ factor with respect to the usual right-handed expressions, in order to obtain the expected positive value for the angular momentum $J_z$ of a rotating geon with $\cos(3t-\phi)$ time and angular dependence. There are six other conformal Killing fields on $\mathscr{I}$, corresponding to space translation and boost on AdS (see \cite{AshMag}), but because of our method of calculating the $l=0$ and $l=1$ spherical harmonic scalar- and vector-type components, the conserved quantities corresponding to them turn out to be vanishing.

\section{Spherical harmonic decomposition} \label{appsphharm}

Complex spherical harmonics are convenient for linear perturbations, but when going to higher order the use of real harmonics makes it much easier to ensure that physical quantities take real values. We use real scalar spherical harmonics that are defined by
\begin{equation}
\mathbb{S}_{lm}=
 \begin{cases}
  (-1)^m\sqrt{\frac{2l+1}{2\pi}\frac{(l-m)!}{(l+m)!}}P_l^m(\cos\theta)\cos(m\phi) & \text{if \ } m>0  \ , \\
  \frac{1}{2}\sqrt{\frac{2l+1}{\pi}}P_l(\cos\theta) & \text{if \ } m=0  \ , \label{eqspharmdef} \\
  (-1)^{|m|}\sqrt{\frac{2l+1}{2\pi}\frac{(l-|m|)!}{(l+|m|)!}}P_l^{|m|}(\cos\theta)\sin(|m|\phi) & \text{if \ } m<0  \ ,
 \end{cases}
\end{equation}
where $P_l$ are Legendre polynomials, and $P_l^m$ are associated Legendre polynomials. The powers of $-1$ in the definition are to cancel the Condon–Shortley phase included in the definition of the associated Legendre polynomials. This way we obtain expressions for $\mathbb{S}_{lm}$ with concrete $l$ and $m$ without alternating signs. On the unit 2-sphere the spherical harmonics satisfy the differential equation
\begin{equation}
D^i D_i \mathbb{S}_{lm}+l(l+1)\mathbb{S}_{lm}=0 \ , \label{slmeq}
\end{equation}
where $D_i$ is the derivative operator belonging to the standard metric $\gamma_{ij}$ on the sphere. The normalization condition is
\begin{equation}
\int_0^\pi {\rm d}\theta\int_0^{2\pi}{\rm d}\phi\sin\theta\,\mathbb{S}_{lm}\mathbb{S}_{
\hat l\hat m}=\delta_{l\hat l}\delta_{m\hat m} \ .
\end{equation}
Using standard spherical coordinates \eqref{slmeq} can be written as
\begin{equation}
\frac{\partial^2\mathbb{S}_{lm}}{\partial\theta^2}
+\frac{\cos\theta}{\sin\theta}\frac{\partial\mathbb{S}_{lm}}{\partial\theta}
+\frac{1}{\sin^2\theta}\frac{\partial^2\mathbb{S}_{lm}}{\partial\phi^2}
+l(l+1)\mathbb{S}_{lm}=0 \ .
\end{equation}
A scalar on the $2$-sphere can be decomposed into spherical harmonic components as
\begin{equation}
f=\sum_{l,m}f_{(lm)}{S}_{lm} \quad , \qquad
f_{(lm)}=\int_0^\pi {\rm d}\theta\int_0^{2\pi}{\rm d}\phi\sin\theta\,\mathbb{S}_{lm}f
\ . \label{scaldecomp}
\end{equation}
The function $f$ may depend on the time and radial coordinates. The $T_{tt}$, $T_{tx}$ and $T_{xx}$ components of a symmetric spacetime tensor $T_{\mu\nu}$ behave as scalars in this respect.

For two-dimensional spheres the vector harmonics $\mathbb{V}_{(lm)i}$ can be expressed in terms of the scalar harmonics as $\mathbb{V}_{(lm)i}=\frac{1}{\sqrt{l(l+1)}}\epsilon_{ij}D^j \mathbb{S}_{lm}$, where $\epsilon_{ij}$ is the natural volume element on the sphere, belonging to the metric $\gamma_{ij}$. Their components in standard angular coordinates are
\begin{equation}
 \mathbb{V}_{(lm)\theta}=\frac{1}{\sqrt{l(l+1)}}
 \frac{1}{\sin\theta}\frac{\partial \mathbb{S}_{lm}}{\partial\phi} \ , \quad
 \mathbb{V}_{(lm)\phi}=\frac{-1}{\sqrt{l(l+1)}}
 \sin\theta\frac{\partial\mathbb{S}_{lm}}{\partial\theta}  \ .
\end{equation}
Clearly, $D^i\mathbb{V}_{(lm)i}=0$. Vector harmonics are only defined for $l\geq1$. They satisfy the same differential equation as the gradient of the scalar harmonics,
\begin{equation}
D^j D_j \mathbb{V}_{(lm)i}+\left[l(l+1)-1\right]\mathbb{V}_{(lm)i}=0 \ ,
\end{equation}
and they are normalized as
\begin{equation}
\int_0^\pi {\rm d}\theta\int_0^{2\pi}{\rm d}\phi\sin\theta\,\mathbb{V}_{(lm)i}\mathbb{V}_{
{(\hat l\hat m)}}^{\ \ \ i}=\delta_{l\hat l}\delta_{m\hat m} \ .
\end{equation}
A covariant vector $v_i$ satisfying $D^iv_i=0$ can be decomposed as
\begin{equation}
v_i=\sum_{l,m}v_{(lm)}\mathbb{V}_{(lm)i}  \quad , \qquad
v_{(lm)}=\int_0^\pi {\rm d}\theta\int_0^{2\pi}{\rm d}\phi\sin\theta\,\mathbb{V}_{(lm)i}v^i
\ . \label{vectdecomp}
\end{equation}

If $v_i=(v_\theta,v_\phi)$ is a general covariant vector on the $2$-sphere, then we can decompose it into its vector-type and scalar-type parts as $v_i=V_i+D_iS$, where $D^iV_i=0$ (see Proposition 2.1 of \cite{IshWal}). To decompose $v_i$, the first thing to do is to calculate the scalar $D^iv_i=D^iD_iS$ and decompose it using \eqref{scaldecomp}. During our concrete calculations, generally there will be only a finite number of nonzero terms. Then \eqref{slmeq} can be used to construct $S$ by adding the $(l,m)$ components of $D^iD_iS$ divided by $-l(l+1)$. After this, the vector-type part $V_i=v_i-D_iS$ can be decomposed in terms of vector harmonics using \eqref{vectdecomp}. The $(T_{t\theta},T_{t\phi})$ and the $(T_{x\theta},T_{x\phi})$ component pairs of a symmetric tensor $T_{\mu\nu}$ behave as covariant vectors in this respect.

The decomposition of a symmetric tensor $T_{ij}$ on the $2$-sphere begins by calculating the trace $T_{ij}\gamma^{ij}$, which is a scalar and can be decomposed according to \eqref{scaldecomp}. Taking the traceless part, $J_{ij}=T_{ij}-T_{k}^{\ k}\gamma_{ij}$, we can define the vector $J_i=D^jJ_{ij}$ and the scalar $J=D^iJ_i$ (see Proposition 2.2 of \cite{IshWal}). The scalar $J$ can be decomposed according to \eqref{scaldecomp}. Defining a scalar $W$ that satisfies $(D^iD_i+2)W=2J$, the function $W$ can be constructed by adding the $(l,m)$ components of $J$ divided by $(1-l)(l+2)/2$. Defining another function, $S$, by $D^iD_iS=W$, we can construct it from the components of $W$, with division by $-l(l+1)$. If we construct the vector $\bar V_i=J_i-\frac{1}{2}D_iW-D_iS$, it has no scalar-type part, so it can be decomposed using \eqref{vectdecomp}. Then $(D^iD_i+1)V_j=2\bar V_j$ defines the the divergence free vector $V_i$, which can be calculated from the components of $\bar V_i$, dividing them by $(1-l)(l+2)/2$. In the end, the decomposition of $T_{ij}$ can be written as
\begin{equation}
T_{ij}=D_{(i}V_{j)}+\left(D_iD_j-\frac{1}{2}\gamma_{ij}D^mD_m\right)S+T_{k}^{\ k}\gamma_{ij} \ ,
\end{equation}
where the first term is the vector-type part, and the rest are scalar-types. On the 2-sphere there are no symmetric tensor spherical harmonics \cite{Higuchi}.

This decomposition procedure of vector and tensors is applied each time when calculating the $(l,m)$ scalar and vector components of the inhomogeneous source terms that arise from lower order perturbations.

\section{Rotations of spherical harmonics} \label{app:rotations}

In this appendix we study the rotational properties of $l=2$ scalar spherical harmonics. According to \eqref{eqspharmdef},
\begin{align}
 \mathbb{S}_{22}&=\frac{1}{4}\sqrt{\frac{15}{\pi}}\sin^2\theta\cos(2\phi) \ , \\
 \mathbb{S}_{21}&=\frac{1}{4}\sqrt{\frac{15}{\pi}}\sin(2\theta)\cos\phi \ , \\
 \mathbb{S}_{20}&=\frac{1}{8}\sqrt{\frac{5}{\pi}}\left[1+3\cos(2\theta)\right] \ , \\
 \mathbb{S}_{2-1}&=\frac{1}{4}\sqrt{\frac{15}{\pi}}\sin(2\theta)\sin\phi \ , \\
 \mathbb{S}_{2-2}&=\frac{1}{4}\sqrt{\frac{15}{\pi}}\sin^2\theta\sin(2\phi) \ .
\end{align}
We define a function by taking a general linear combination of these,
\begin{equation}
 F=\sum_{m=-2}^{2}\alpha_m\mathbb{S}_{2m} \ ,
\end{equation}
and make a rotation with Euler angles $\alpha$, $\beta$, and $\gamma$. We use the convention that first a rotation with angle $\alpha$ is made around the $z$ axis, then a rotation $\beta$ around the new $x$ axis, and finally a rotation $\gamma$ around the new $z$ axis. The relation between the coordinates on the unit sphere is $z=\cos\theta$, $x=\sin\theta\cos\phi$, and $y=\sin\theta\sin\phi$. We decompose the rotated function in terms of the original spherical harmonics,
\begin{equation}
 \bar F=\sum_{m=-2}^{2}\bar\alpha_m\mathbb{S}_{2m} \ .
\end{equation}
The rotated spherical harmonic components can be written as
\begin{align}
 \bar\alpha_{-2}&=A_1\cos(2\gamma)+A_2\sin(2\gamma) \ , \\
 \bar\alpha_{-1}&=A_3\cos\gamma+A_4\sin\gamma \ , \label{eqalm1} \\
 \bar\alpha_{0}&=-\sqrt{3}\left[B_3\sin(2\beta)-B_4\cos(2\beta)-\frac{B_5}{3}\right] \ , \\
 \bar\alpha_{1}&=A_3\sin\gamma-A_4\cos\gamma \ , \label{eqalp1} \\
 \bar\alpha_{2}&=A_1\sin(2\gamma)-A_2\cos(2\gamma) \ ,
\end{align}
where
\begin{align}
 A_1&=B_1\cos\beta+B_2\sin\beta \ , \\
 A_2&=B_3\sin(2\beta)-B_4\cos(2\beta)+B_5 \ , \\
 A_3&=2B_3\cos(2\beta)+2B_4\sin(2\beta) \ , \\
 A_4&=B_1\sin\beta-B_2\cos\beta \ , \\
 B_1&=\alpha_{-2}\cos(2\alpha)-\alpha_{2}\sin(2\alpha) \ , \\
 B_2&=\alpha_{1}\cos\alpha+\alpha_{-1}\sin\alpha \ , \\
 B_3&=\frac{1}{2}\left(\alpha_{-1}\cos\alpha-\alpha_{1}\sin\alpha\right) , \\
 B_4&=\frac{1}{4}\left(\sqrt{3}\alpha_{0}+\alpha_{2}\cos(2\alpha)
  +\alpha_{-2}\sin(2\alpha)\right) , \\
 B_5&=\frac{1}{4}\left(\sqrt{3}\alpha_{0}-3\alpha_{2}\cos(2\alpha)
  -3\alpha_{-2}\sin(2\alpha)\right) . \label{eqb5rotapp}
\end{align}
We note that $\mathbb{S}_{22}$ can be rotated into $\mathbb{S}_{21}$ by an Euler rotation with $\alpha=\pi/4$, $\beta=\pi/2$, and $\gamma=0$.

It is expected that by an Euler rotation one can make three of the five $\bar\alpha_m$ zero. However, this is not true for any three of them. We prove the claim stated in Sec.~\ref{seclofrgeonlin}, that it is always possible to make $\bar\alpha_{-2}=\bar\alpha_{-1}=\bar\alpha_{1}=0$. It is obviously enough to prove that one can make $\bar\alpha_{-1}=\bar\alpha_{1}=0$, since after that it is easy to make $\bar\alpha_{-2}=0$ by a simple rotation with some angle $\gamma$ around the $z$ axis. It follows from \eqref{eqalm1} and \eqref{eqalp1} that $\bar\alpha_{-1}=\bar\alpha_{1}=0$ if and only if $A_3=A_4=0$. From $A_3=0$ it is possible to express $\tan(2\beta)$, and $A_4=0$ can be solved for $\tan\beta$. Using the identity $\tan(2\beta)=2\tan\beta/(1-\tan^2\beta)$ we get an equation involving only $\alpha$. Using the identity $\sin^2\alpha=2\tan\alpha/(1+\tan^2\alpha)$, this equation can be reduced to a third order polynomial equation in $\tan\alpha$, which always has at least one real root.

\section{Nonlinear gauge-invariant formalism} \label{apphighordg}

The higher order generalization of the linear gauge-invariant perturbation formalism of \cite{KodIshSet,KodIsh2003} was used in \cite{Dias12a,Horowitz14,Dias16a,Dias17a} for the construction of AdS geon solutions. In this appendix we give more details about the nonlinear generalization of the Kodama-Ishibashi-Seto method. We do not attempt to present a general formalism that is gauge invariant to several orders at the same time. Our aim here is to clarify the transformation of the variables for a given order gauge change and to motivate the choice of gauge and the used variables in Appendix \ref{app:vectpert} and \ref{app:scalpert} and in the main part of the manuscript. For the concrete calculation of geon configurations we make a specific gauge choice at each order, so a detailed gauge-invariant formalism is not crucial for us.

At each order in the $\varepsilon$ expansion there is a gauge freedom to choose what coordinate system we use. In the main part of this paper, at zeroth order we use the metric form \eqref{sphttx} with coordinates $t$ and $x$. Then we can make an $\varepsilon^k$ order gauge transformation for each $k\geq 1$ in increasing order.

Let us take an $\varepsilon$ dependent coordinate transformation from coordinates $x^\mu$ to $\bar x^\mu$. We assume that it is a $k$th order transformation generated by a vector field $\xi^\mu$, and the inverse transformation can be expanded as
\begin{equation}
 x^\mu(\bar x,\varepsilon)=\bar x^\mu+\varepsilon^k\xi^\mu(\bar x)+\mathcal{O}(\varepsilon^{k+1}) \ .
\end{equation}
There should be an index $(k)$ denoting the order on $\xi^\mu$, but we drop it to simplify the formulas. In this case the $k$th order metric perturbation in expansion \eqref{gmunuexp1} transforms as
\begin{equation}
 \bar g_{\mu\nu}^{(k)}=g_{\mu\nu}^{(k)}+\nabla_\mu^{(0)}\xi_{\nu}+\nabla_\nu^{(0)}\xi_{\mu} \ ,
\end{equation}
where $\nabla_\mu^{(0)}$ is the derivative operator belonging to the AdS background metric $g_{\mu\nu}^{(0)}$.

\subsection{Vector-type perturbation}

Using the notation of \cite{IshWal}, the components of $\varepsilon^k$ order vector-type metric perturbations in the $\mathbb{V}_{(lm)i}$ class can be written as
\begin{equation}
g^{(k)}_{ab}=0 \ , \quad g^{(k)}_{ai}=H^{(v)}_{a}\mathbb{V}_i \ , 
\quad g^{(k)}_{ij}=2H^{(v)}_T D_{(i}\mathbb{V}_{j)} \ , \label{genmetvectpert}
\end{equation}
where from $H^{(v)}_{a}$, $H^{(v)}_T$, and $\mathbb{V}_i$ we have dropped the $lm$ indices, and from $H^{(v)}_{a}$ and $H^{(v)}_T$ the reference that they are order $k$ quantities. Here, as in \eqref{eqbackgrm}, the indices $a,b$ correspond to coordinates in the time-radius plane, $y^a=(t,x)$, and $i,j$ correspond to coordinates in the standard 2-sphere, on which the covariant derivative is denoted by $D_i$. General $k$th order vector perturbations can be written as linear combinations of these terms for all possible $l$ and $m$. The functions $H^{(v)}_{a}$ and $H^{(v)}_T$ depend only on the coordinates $y^a$. If $l=1$ then $D_{(i}\mathbb{V}_{j)}=0$, so $\mathbb{V}_{i}$ is a Killing vector field. Then $g^{(k)}_{ij}=0$, and $H^{(v)}_T$ is not defined. 

The most general vector-type gauge transformation that keeps $g^{(k)}_{\mu\nu}$ in the vector $\mathbb{V}_{i(lm)}$ class is generated by a vector $\xi^\mu$, for which $\xi_a=0$ and $\xi_i=\xi^{(v)}\mathbb{V}_{i}$, where $\xi^{(v)}$ is a scalar function on the $y^a$ plane. It can be checked easily that the metric perturbation functions transform as
\begin{align}
H^{(v)}_{a}&\to H^{(v)}_{a}-r^2\hat\nabla_a\left(\frac{\xi^{(v)}}{r^2}\right) \ , \label{hatrans} \\
H^{(v)}_{T}&\to H^{(v)}_{T}-\xi^{(v)} \qquad \text{for} \quad l\geq 2 \  , \label{httrans}
\end{align}
where $\hat\nabla_a$ is the derivative operator on the $y^a$ plane of the AdS background metric. 

For $l\geq 2$ the function $H^{(v)}_{T}$ is always defined, and the combination
\begin{equation}
Z_a=H^{(v)}_{a}-r^2\hat\nabla_a\left(\frac{H^{(v)}_{T}}{r^2}\right) \label{zinvdef}
\end{equation}
is independent of $\xi^{(v)}$, so it is gauge invariant. It is the vector-type Kodama-Ishibashi gauge-invariant variable\cite{KodIshSet}. A natural way to proceed with the actual calculations for $l\geq 2$ is to use the gauge freedom in \eqref{httrans} to make $H^{(v)}_{T}=0$, in which case $Z_a=H^{(v)}_{a}$ and $g^{(k)}_{ij}=0$. This choice eliminates the highest angular derivatives from \eqref{genmetvectpert}, and corresponds to the Regge-Wheeler gauge in the literature \cite{Nollert1999}.

For the $l=1$ case the only gauge-invariant quantity is
\begin{equation}
Z_{ab}=r^2\hat\nabla_a\left(\frac{H^{(v)}_{b}}{r^2}\right)
-r^2\hat\nabla_b\left(\frac{H^{(v)}_{a}}{r^2}\right) \ , \label{zabvectpert}
\end{equation}
which has only one independent component. In the $(t,x)$ coordinate system,
\begin{equation}
 Z_{tx}=
 \frac{\partial H^{(v)}_{x}}{\partial t}
 -\frac{\partial H^{(v)}_{t}}{\partial x}
 +\frac{2}{\sin x\cos x}H^{(v)}_{t} \ . \label{ztxeqapp}
\end{equation}

\subsection{Scalar-type perturbation}

The components of $\varepsilon^k$ order scalar-type metric perturbations in the $\mathbb{S}_{lm}$ class can be written as
\begin{align}
g^{(k)}_{ab}&=H^{(s)}_{ab}\mathbb{S} \quad , \qquad 
g^{(k)}_{ai}=H^{(s)}_{a}D_i\mathbb{S} \ , \notag \\
g^{(k)}_{ij}&=H^{(s)}_L\gamma_{ij}\mathbb{S}
+H^{(s)}_T\left(D_i D_j+\frac{l(l+1)}{2}\gamma_{ij}\right)\mathbb{S} \ , \label{gabgaigijgen}
\end{align}
where the functions $H^{(s)}_{ab}$, $H^{(s)}_{a}$, $H^{(s)}_L$ and $H^{(s)}_T$ depend only on $y^a$. We have dropped the $lm$ indices and the reference that we are at order $k$ from the functions. The general scalar perturbation is the linear combination of these for all possible $l$ and $m$. If $l=0$ then $D_i\mathbb{S}=0$, $g^{(k)}_{ai}=0$, and $H^{(s)}_{a}$ is not defined. If $l=0$ or $l=1$ then $\left(D_i D_j+\frac{l(l+1)}{2}\gamma_{ij}\right)\mathbb{S}=0$, in which cases $H^{(s)}_T$ cannot be defined.

The most general $k$th order coordinate transformations that keep $g^{(k)}_{\mu\nu}$ in the scalar $\mathbb{S}_{lm}$ class are generated by vector fields $\xi^\mu$ that has the components
\begin{equation}
\xi_{a}=\xi^{(s)}_a \mathbb{S} \ , \quad \xi_{i}=\xi^{(s)} D_i\mathbb{S} \ ,
\end{equation}
where $\xi^{(s)}_a$ and $\xi^{(s)}$ depend on the coordinates $y^a$, and $\xi^{(s)}$ is defined only for $l\geq1$. The metric perturbation functions transform as
\begin{align}
H^{(s)}_{ab}&\to H^{(s)}_{ab}
-\hat\nabla_a\xi^{(s)}_b
-\hat\nabla_b\xi^{(s)}_a \ , \label{habtrans} \\
H^{(s)}_{L}&\to H^{(s)}_{L}
+l(l+1)\xi^{(s)}-2r(\hat\nabla^a r)\xi^{(s)}_a \ , \label{hl0trans} \\
H^{(s)}_{a}&\to H^{(s)}_{a}
-\xi^{(s)}_a
-r^2\hat\nabla_a\left(\frac{\xi^{(s)}}{r^2}\right) \quad \text{for} \quad l\geq 1  \ ,  \label{ha0trans} \\
H^{(s)}_{T}&\to H^{(s)}_{T}-2\xi^{(s)} \quad \text{for} \quad l\geq 2  \ . \label{ht0trans}
\end{align}
For the $l=0$ case we have to drop the $l(l+1)\xi^{(s)}$ term from \eqref{hl0trans}. In terms of the $(t,x)$ coordinates,
\begin{align}
 H^{(s)}_{tt}&\to H^{(s)}_{tt}
 -2\frac{\partial\xi^{(s)}_{t}}{\partial t}+2\tan x\xi^{(s)}_x \ ,\\
 H^{(s)}_{tx}&\to H^{(s)}_{tx}
 -\frac{\partial\xi^{(s)}_{t}}{\partial x}
 -\frac{\partial\xi^{(s)}_x}{\partial t}
 +2\tan x\xi^{(s)}_{t} \ , \label{htxtrans} \\
 H^{(s)}_{xx}&\to H^{(s)}_{xx}
 -2\frac{\partial\xi^{(s)}_x}{\partial x}+2\tan x\xi^{(s)}_x \ ,\\
 H^{(s)}_L&\to H^{(s)}_L+l(l+1)\xi^{(s)}
 -2\tan x\xi^{(s)}_x \ . \label{hl0transtx}
\end{align}

It is easy to check that for $l\geq 2$ the scalar-type Kodama-Ishibashi-Seto gauge-invariant variables 
\begin{align}
Z&=\frac{2}{r^2}\left[H^{(s)}_{L}+\frac{l(l+1)}{2}H^{(s)}_{T}
+2r(\hat\nabla^a r)X_a\right] , \\
Z_{ab}&=H^{(s)}_{ab}
+\hat\nabla_a X_b+\hat\nabla_b X_a
+\frac{1}{2}Z\hat g_{ab} \ ,
\end{align}
are gauge invariant\cite{KodIshSet}, where $X_a$ is defined by
\begin{equation}
X_a=-H^{(s)}_{a}
+\frac{r^2}{2}\hat\nabla_a\left(\frac{H^{(s)}_T}{r^2}\right) \ .
\end{equation}
For our calculations we use the freedom \eqref{ha0trans} to set $H^{(s)}_{a}=0$ by choosing $\xi^{(s)}_a$, and we use \eqref{ht0trans} to set $H^{(s)}_T=0$ by choosing $\xi^{(s)}$. In this gauge $X_a=0$, and the Kodama-Ishibashi gauge-invariant variables are simply
\begin{equation}
Z=\frac{2}{r^2}H^{(s)}_L \quad , \qquad
Z_{ab}=H^{(s)}_{ab}+\frac{1}{r^2}H^{(s)}_L\hat g_{ab} \ . \label{zzingauge}
\end{equation}
This choice is usually called Regge-Wheeler gauge in the literature. In this gauge, the metric perturbation does not have $g^{(n)}_{ai}$ components, and $g^{(n)}_{ij}$ is proportional to $\gamma_{ij}$, in particular, $g^{(n)}_{\theta\phi}=0$. 

If $l=0$ then $H^{(s)}_a$ and $H^{(s)}_T$ are not defined. A natural choice is to use the freedom in \eqref{hl0transtx} to make $H^{(s)}_L=0$ by choosing $\xi^{(s)}_x$ appropriately. We also make a choice for $\xi^{(s)}_{t}$ in \eqref{htxtrans} which makes $H^{(s)}_{tx}=0$. Even after this, we have the freedom to make an additional gauge transformation with
\begin{equation}
\xi^{(s)}_{t}=\frac{f_1(t)}{\cos^2 x} \ , \label{eqxisttrans}
\end{equation}
where $f_1(t)$ is an arbitrary function, because this leaves $H^{(s)}_{tx}$ unchanged. Since this transformation is generated by the vector with components $\xi^{(s)t}=-f_1(t)/L^2$ and $\xi^{(s)x}=0$, the residual freedom corresponds to the relabeling of the constant time hypersurfaces.

\section{Vector-type perturbations} \label{app:vectpert}

We consider $\varepsilon^k$ order class $\mathbb{V}_{(lm)i}$ vector-type perturbations of the metric tensor in the Regge-Wheeler gauge where the $g^{(k)}_{ij}$ metric components are vanishing ($a,b=1,2$ and $i,j=3,4$),
\begin{equation}
 g^{(k)}_{ab}=0 \ , \quad g^{(k)}_{ai}=H^{(v)}_{a}\mathbb{V}_i \ ,
 \quad g^{(k)}_{ij}=0 \ , \label{vectgauge}
\end{equation}
where $H^{(v)}_{a}$ depend only on the coordinates $y^a=(t,x)$. We have dropped the $lm$ indices from the functions, and also the reference that we are at $\varepsilon^k$ order. The gauge invariant formalism described in Appendix \ref{apphighordg} shows that one can always make this gauge choice. The method we present here is based on the linear formalism presented in \cite{IshWal}.

\subsection{Case \texorpdfstring{$l\geq 2$}{l>=2}}

If $l\geq 2$, the Kodama-Ishibashi gauge-invariant variables $Z_a$ can be defined, and in our gauge
\begin{equation}
 Z_a=H^{(v)}_{a}  \label{eqzhgauge}
\end{equation}
(see Eq.~\eqref{zinvdef}). For simplicity, in the Regge-Wheeler gauge this equation can be considered as the definition of $Z_a$.

For perturbations of the form \eqref{vectgauge} the $(a,b)$ components of the $\varepsilon^k$ order Einstein equations are identically satisfied. All $(i,j)$ components give the same condition,
\begin{equation}
\frac{\partial Z_t}{\partial t}-\frac{\partial Z_x}{\partial x}=T \ , \label{zttzxxcond}
\end{equation}
where $T$ is a source term arising from lower order perturbations, which is assumed to be already known and calculated. Let us denote the $t$ independent part of the source term by $T_1$. The time dependent part, $T_2=T-T_1$, generally has the $t$ dependence through $\sin(\omega t)$ and $\cos(\omega t)$ terms, where the frequency $\omega$ is some integer. The equation $\frac{\partial}{\partial t}\bar Z_t=T_2$ can be integrated easily to obtain the function $\bar Z_t$. The time independent part $T_1$ turns out to be zero in most of the cases, but even if not, the equation $-\frac{\partial}{\partial x}\bar Z_x=T_1$ can be solved for $\bar Z_x$. The general solution of \eqref{zttzxxcond} can be written in terms of an arbitrary scalar function $\phi_V$, 
\begin{equation}
Z_t=\frac{\partial\phi_V}{\partial x}+\bar Z_t \quad , \qquad
Z_x=\frac{\partial\phi_V}{\partial t}+\bar Z_x \ . \label{eqsztzx}
\end{equation}
It is a general principle that during our computations it is practical to avoid calculating integrals in $x$, since performing integrals with respect to $t$ is generally much simpler because of the simple trigonometric time dependence.

After making the substitution for $Z_a$, it turns out that the $(a,i)$ components of the Einstein equations are not independent. The $(x,i)$ and $(t,i)$ components of the Einstein equations are equivalent to
\begin{equation}
 \frac{\partial E_V}{\partial t}=0 \quad , \qquad
 \frac{\partial E_V}{\partial x}=0 \ , \label{devtdevxeq}
\end{equation}
respectively, where 
\begin{equation}
E_V=\sin^2 x\left(\frac{\partial^2\phi_V}{\partial x^2}-\frac{\partial^2\phi_V}{\partial t^2}\right)
-2\tan x\frac{\partial\phi_V}{\partial x}-(l+2)(l-1)\phi_V+\phi_V^{(0)} \ . \label{evexpression}
\end{equation}
The function $\phi_V^{(0)}$ is determined by the lower order perturbations. 

A concrete expression for $\phi_V^{(0)}$ can be obtained by comparing the derivatives of \eqref{evexpression} to the actual form of the field equations obtained by the algebraic manipulation software. As the result of the comparison, we have to solve two equations,
\begin{equation}
 \frac{\phi_V^{(0)}}{\partial t}=T_{t} \quad , \qquad
 \frac{\phi_V^{(0)}}{\partial x}=T_{x} \ , \label{eqphiv0dtdx}
\end{equation}
where $T_a$ are given functions of $t$ and $x$, satisfying the necessary integrability condition. The solution for $\phi_V^{(0)}$ can be obtained in a straightforward way, but generally it involves the calculation of an integral in $x$, which can be hard to obtain in closed form at higher orders in the $\varepsilon$ expansion. For example, at $\varepsilon^5$ order, polylogarithm functions tend to arise.

It is possible to avoid the calculation of integrals when solving \eqref{eqphiv0dtdx}, since generally the source terms are sums of a finite number of Fourier components with integer frequencies,
\begin{equation}
 T_{a}=\sum_{\bar k=0}^{\bar k_{\mathrm{max}}}
 \left(T_{a}^{(c,\bar k)}\cos(\bar k t)+T_{a}^{(s,\bar k)}\sin(\bar k t)\right) ,
\end{equation}
and we can search for the solution in a similar expansion form,
\begin{equation}
 \phi_V^{(0)}=\sum_{\bar k=0}^{\bar k_{\mathrm{max}}}\left(
 \phi_V^{(0,c,\bar k)}\cos(\bar k t)+\phi_V^{(0,s,\bar k)}\sin(\bar k t)
 \right) . \label{eqphiv0exp}
\end{equation}
If $\bar k>0$, then it is possible to solve the first equation of \eqref{eqphiv0dtdx} algebraically for $\phi_V^{(0,c,\bar k)}$ and $\phi_V^{(0,s,\bar k)}$. However, the $\bar k=0$ component of the first equation is satisfied identically, and the second gives an expression for the $x$ derivative of $\phi_V^{(0,c,0)}$. Luckily, it is not necessary to calculate the integral to get the concrete form of $\phi_V^{(0,c,0)}$, since the calculation of the gauge invariant variables $Z_a$ in \eqref{eqsztzx} only requires the differentiated form of the zeroth Fourier component of $\phi_V$.

It follows from \eqref{devtdevxeq} that $E_V$ is a constant. However, since we can add a constant to $\phi_V$ without changing $Z_t$ and $Z_x$ in \eqref{eqsztzx}, we can shift this constant to zero, setting $E_V=0$. We define a rescaled scalar function by
\begin{equation}
\phi_V=r\Phi_V \ ,
\end{equation}
where $r=L\tan x$. Then the $E_V=0$ condition can be written into the simple form
\begin{equation}
-\frac{\partial^2\Phi_V}{\partial t^2}+\frac{\partial^2\Phi_V}{\partial x^2}
-\frac{l(l+1)}{\sin^2 x}\Phi_V+\frac{\Phi_V^{(0)}}{\sin^2 x}=0 \ , \label{eqmastvect}
\end{equation}
where
\begin{equation}
 \Phi_V^{(0)}=\frac{1}{r}\phi_V^{(0)} \ . \label{eqbigsmallphiv}
\end{equation}
Equation \eqref{eqmastvect} is the master equation describing all $l\geq 2$ vector-type perturbations. We show in Appendix \ref{app:scalpert} that $l\geq 2$ scalar-type perturbations are described by an equation having exactly the same form, although the boundary conditions at infinity are different in the two cases.

We give a detailed description of how to solve the inhomogeneous differential equation \eqref{eqmastvect} in Appendix \ref{apptimeper}. After the solution for $\Phi_V$ with the appropriate boundary condition is obtained in the way described there, then from \eqref{eqbigsmallphiv} we get $\phi_V$, and using \eqref{eqsztzx} we obtain $Z_a$. According to \eqref{eqzhgauge} in our gauge $Z_a=H^{(v)}_{a}$, and the resulting class $\mathbb{V}_{(lm)i}$ vector-type metric perturbation components can be obtained by \eqref{vectgauge}.

\subsection{Case \texorpdfstring{$l=1$}{l=1}}

In this case the first step is to consider the value of
\begin{equation}
 Z_{tx}=
 \frac{\partial H^{(v)}_{x}}{\partial t}
 -\frac{\partial H^{(v)}_{t}}{\partial x}
 +\frac{2}{\sin x\cos x}H^{(v)}_{t} \ , \label{ztxeqsec}
\end{equation}
which is the only nonzero component of the antisymmetric $Z_{ab}$ gauge invariant quantity given in \eqref{zabvectpert}. For simplicity, \eqref{ztxeqsec} may also be considered as the definition of $Z_{tx}$. It follows from the $(x,\phi)$ component of the Einstein equations that the time derivative of $Z_{tx}$ is zero. The $(t,\phi)$ component gives a time-independent value for the $x$ derivative of $\bar Z_{tx}=Z_{tx}\sin^2 x$. There are no further restrictions from the remaining components of the field equations. The expression for $\bar Z_{tx}$ can be calculated by integration, and the regularity of the corresponding metric perturbation at the center can be ensured by setting $\bar Z_{tx}=0$ at $x=0$.

Defining $\bar H^{(v)}_{t}=r^{-2}H^{(v)}_{t}$ and $\bar H^{(v)}_x=r^{-2}H^{(v)}_x$, Eq.~\eqref{ztxeqsec} can be written as
\begin{equation}
 \frac{Z_{tx}}{r^2}=\frac{\partial\bar H^{(v)}_{x}}{\partial t}
 -\frac{\partial\bar H^{(v)}_{t}}{\partial x}\ .
\end{equation}
Here $Z_{tx}$ is already known, and apart from this equation there are no restrictions on the metric variables $H^{(v)}_{a}$. A particular time independent solution can be found by setting $\bar H^{(v)}_x=0$ and integrating from $x=\pi/2$ to obtain $\bar H^{(v)}_{t}=\bar H^{(v,0)}_{t}$. Then the general solution is
\begin{equation}
 \bar H^{(v)}_{t}=\bar H^{(v,0)}_{t}+\frac{\partial f}{\partial t} \quad \ , \qquad
 \bar H^{(v)}_x=\frac{\partial f}{\partial x} \ ,
\end{equation}
where $f$ is an arbitrary function of $t$ and $x$. The function $f$ corresponds to the gauge freedom $\xi^{(v)}/r^2$ in \eqref{hatrans}. If $f$ would depend on $t$ at $x=\pi/2$ then $H^{(v)}_{t}$, and consequently $g^{(k)}_{t\phi}$, would diverge as $(\pi/2-x)^{-2}$. This would correspond to an asymptotically rotating coordinate system. When the spherical harmonic index $m$ is zero, the freedom in $f$ corresponds to a time and radius dependent rotation in the $\phi$ direction. In our calculations we make the natural gauge choice $f=0$. Then $H^{(v)}_{t}=r^2\bar H^{(v,0)}_{t}$ and $H^{(v)}_{x}=0$, and the metric perturbation is necessarily regular and asymptotically AdS.

\section{Scalar-type perturbations} \label{app:scalpert}

We consider $\varepsilon^k$ order scalar-type perturbations in the class belonging to the $\mathbb{S}_{lm}$ real spherical harmonics in the Regge-Wheeler gauge, where the metric perturbation is in the block diagonal form ($a,b=1,2$ and $i,j=3,4$)
\begin{equation}
g^{(k)}_{ab}=H^{(s)}_{ab}\mathbb{S} \ , \quad 
g^{(k)}_{ai}=0 \ , \quad 
g^{(k)}_{ij}=H^{(s)}_L\gamma_{ij}\mathbb{S} \ . \label{habhaihij}
\end{equation}
Here the functions $H^{(s)}_{ab}$ and $H^{(s)}_L$ depend only on the coordinates $y^a=(t,x)$, and $\gamma_{ij}$ is the standard metric on the 2-sphere with coordinates $z^i=(\theta,\phi)$. We have dropped the $lm$ indices from all these functions, and also the reference that we are at $\varepsilon^k$ order. That we can make this gauge choice follows from the considerations in Appendix \ref{apphighordg}. The formalism presented here is based on the work in \cite{IshWal} on linear perturbations.

\subsection{Case \texorpdfstring{$l\geq 2$}{l>=2}}

If $l\geq 2$, the Kodama-Ishibashi-Seto gauge-invariant variables $Z$ and $Z_{ab}$ can be defined\cite{KodIshSet}, and in the Regge-Wheeler gauge the metric variables can be expressed as (see Eq.~\eqref{zzingauge})
\begin{equation}
H^{(s)}_L=\frac{r^2}{2}Z \quad , \qquad
H^{(s)}_{ab}=Z_{ab}-\frac{1}{2}Z\hat g_{ab} \ . \label{hl0hab0}
\end{equation}
Here $\hat g_{ab}$ is the metric induced on the time-radius plane by the background AdS metric. For simplicity, in the following considerations \eqref{hl0hab0} can be considered as the definition of $Z$ and $Z_{ab}$.

The part of the angular components of the $\varepsilon^k$ order Einstein's equations that is not proportional to $\gamma_{ij}$ gives only one condition, $Z=\hat g^{ab}Z_{ab}+\bar Z$, which in our coordinate system can be written as
\begin{equation}
Z=\frac{\cos^2 x}{L^2}\left(Z_{xx}-Z_{tt}\right)+\bar Z , \label{zeqzab}
\end{equation}
where $\bar Z$ is an inhomogeneous source term depending on $t$ and $x$, fixed by the lower order perturbations.

After substituting $Z$ from \eqref{zeqzab}, the $(a,i)$ components of the field equations are equivalent to
\begin{equation}
\hat\nabla_b Z^b_{\ a}-\hat\nabla_a Z^c_{\ c}=S_a \ , \label{ztenseq}
\end{equation}
where $S_{a}$ is a known source term, depending on the coordinates $t$ and $x$, arising from lower order perturbations, and $\hat\nabla_a$ is the derivative operator in the two-dimensional time-radius plane of the AdS background. Our first task is to find a particular solution $Z_{ab}=\bar Z_{ab}$ of \eqref{ztenseq}. Then the general solution can be written as $Z_{ab}=\hat Z_{ab}+\bar Z_{ab}$, where $\hat Z_{ab}$ is a solution of the homogeneous part of the equation, 
\begin{equation}
\hat\nabla_b \hat Z^b_{\ a}-\hat\nabla_a \hat Z^c_{\ c}=0 .  \label{eqhomzab}
\end{equation}
The expression for $\bar Z_{ab}$ is not unique, and one may choose any solution that is technically easy to obtain.

Using our coordinate system, the $(t,\theta)$ and $(t,\phi)$ components of the field equations are both equivalent to
\begin{equation}
\frac{\partial Z_{tx}}{\partial x}-\frac{\partial Z_{xx}}{\partial t}
=\frac{L^2}{\cos^2 x}S_{t} \ ,  \label{zabeq1}
\end{equation}
and the $(x,\theta)$ and $(x,\phi)$ components are 
\begin{equation}
\frac{\partial Z_{tt}}{\partial x}-\frac{\partial Z_{tx}}{\partial t}
+\tan x(Z_{xx}-Z_{tt})
=\frac{L^2}{\cos^2 x}S_x \ .  \label{zabeq2}
\end{equation}

Similar to the method used for \eqref{zttzxxcond}, Eq.~\eqref{zabeq1} can be solved by decomposing $S_{t}$ into a time independent part $S_{t}^{(1)}$, and to the rest with periodically oscillating time dependence, $S_{t}^{(2)}=S_{t}-S_{t}^{(1)}$. Then integrating
\begin{equation}
\frac{\partial Z_{tx}^{(0)}}{\partial x}
=\frac{L^2}{\cos^2 x}S_{t}^{(1)} \qquad , \quad
-\frac{\partial Z_{xx}^{(0)}}{\partial t}
=\frac{L^2}{\cos^2 x}S_{t}^{(2)}
\end{equation}
gives a particular solution. The general solution of \eqref{zabeq1} can be written as
\begin{equation}
 Z_{tx}=\frac{\partial f_1}{\partial t}+Z_{tx}^{(0)} \ , \qquad 
 Z_{xx}=\frac{\partial f_1}{\partial x}+Z_{xx}^{(0)} \ ,  \label{ztxzxxeqs}
\end{equation}
where $f_1$ is an arbitrary function of $t$ and $x$. 

Substituting \eqref{ztxzxxeqs} into \eqref{zabeq2}, and choosing 
\begin{equation}
 f_1=-\cot x Z_{tt} \ , \label{eqf1choice}
\end{equation}
the $x$ derivatives drop out, and we get
\begin{equation}
 \frac{\partial^2 Z_{tt}}{\partial t^2}+Z_{tt}=f_2 \ , \label{eqzttf2}
\end{equation}
where
\begin{equation}
 f_2=\frac{L\sin x}{\cos^3 x}S_x-\tan^2 x Z_{xx}^{(0)}
\end{equation}
is a known function of $t$ and $x$. Equation \eqref{eqzttf2} can be solved for $Z_{tt}$, again without integrating in $x$,
\begin{equation}
 Z_{tt}=f_c\cos t+f_s\sin t+\sin t\int f_2\cos t {\rm d} t- \cos t\int f_2\sin t {\rm d} t \ ,
\end{equation}
where $f_c$ and $f_s$ are arbitrary functions of $x$. Choosing $f_c$ and $f_s$ appropriately, we get a particular solution $Z_{tt}=\bar Z_{tt}$, which has no terms with exactly $\cos t$ or $\sin t$ time dependence. Substituting \eqref{eqf1choice} into \eqref{ztxzxxeqs} we also get particular solutions for the other components, $Z_{tx}=\bar Z_{tx}$ and $Z_{xx}=\bar Z_{xx}$.

According to Eq.~(117) of \cite{IshWal}, the general solution of the homogeneous equation  \eqref{eqhomzab} can be generated by a function $\phi_S$ as
\begin{equation}
\hat Z_{ab}=\left(\hat\nabla_a\hat\nabla_b
-\frac{1}{L^2}\hat g_{ab}\right)\phi_S \ . \label{eqzabphis}
\end{equation}
Using the $(t,x)$ coordinates, this has the components
\begin{align}
\hat Z_{tt}&=\frac{\partial^2\phi_S}{\partial t^2}-\tan x\frac{\partial\phi_S}{\partial x} 
+\frac{\phi_S}{\cos^2 x} \ , \label{zttphis}\\
\hat Z_{tx}&=\frac{\partial^2\phi_S}{\partial t\partial x}-\tan x\frac{\partial\phi_S}{\partial t} \ , \\
\hat Z_{xx}&=\frac{\partial^2\phi_S}{\partial x^2}-\tan x\frac{\partial\phi_S}{\partial x} 
-\frac{\phi_S}{\cos^2 x} \ . \label{zxxphis}
\end{align}
When we already have both $\bar Z_{ab}$ and $\hat Z_{ab}$, the general solution of the inhomogeneous equation \eqref{ztenseq} can be written as
\begin{equation}
 Z_{ab}=\hat Z_{ab}+\bar Z_{ab} \ . \label{eqzabsol}
\end{equation}
The part proportional to $\gamma_{ij}$ of the angular part of the field equations can be checked to be satisfied at this stage.

In terms of a rescaled scalar function $\phi_S$ defined by
\begin{equation}
\phi_S=r\Phi_S  \ , \label{lowupphis}
\end{equation}
the expressions for $\hat Z_{ab}$ in \eqref{zttphis}-\eqref{zxxphis} can be written into the alternative form
\begin{align}
 \hat Z_{tt}&=L\tan x\left(\frac{\partial^2\Phi_S}{\partial t^2}
 -\tan x\frac{\partial\Phi_S}{\partial x}\right) , \label{zttphisb}\\
 \hat Z_{tx}&=L\left(\tan x\frac{\partial^2\Phi_S}{\partial t\partial x}
 +\frac{\partial\Phi_S}{\partial t}\right) , \\
 \hat Z_{xx}&=L\tan x\frac{\partial^2\Phi_S}{\partial x^2}
 +L\left(1+\frac{1}{\cos^2 x}\right)\frac{\partial\Phi_S}{\partial x} \ . \label{zxxphisb}
\end{align}

The asymptotic behavior of the previously obtained particular solution $\bar Z_{ab}$ is generally not yet ideal for our purposes. The limit
\begin{equation}
 Z_l=L^2\lim_{x\to\pi/2}\left(\Omega^2\bar Z_{tt}\right)
 =-L^2\lim_{x\to\pi/2}\left(\Omega^2\bar Z_{xx}\right) ,
\end{equation}
where $\Omega$ is the conformal factor defined in \eqref{eqconffact}, is generally a nonvanishing function of $t$. These types of divergent terms would make the treatment of asymptotically AdS spacetimes more difficult, since they would affect the metric induced on the conformal boundary. Luckily, we can easily cancel these second-order divergent terms in $\bar Z_{ab}$ by redefining $\phi_S$ into $\phi_S-Z_l$ and absorbing the new terms into $\bar Z_{ab}$ in \eqref{eqzabsol}. This way, the time dependent $Z_l/\cos^2x$ terms in \eqref{zttphis} and \eqref{zxxphis} will exactly cancel the second-order divergent terms in $\bar Z_{ab}$. In the following we assume that we have already achieved $Z_l=0$, which will also make simpler the asymptotically AdS condition on the still unknown function $\phi_S$.

The function $\phi_S$ will be determined by the $(a,b)$ components of the Einstein equations. It can be checked, that similar to the first order homogeneous case considered in \cite{IshWal}, after substituting $Z$ from \eqref{zeqzab} and $Z_{ab}$ from \eqref{eqzabsol}, the $(a,b)$ components are not independent. They can be written in terms of a single scalar function $E_S$,
\begin{equation}
\left(\hat\nabla_a\hat\nabla_b
-\frac{1}{L^2}\hat g_{ab}\right)E_S=0 \ . \label{diffeqephis}
\end{equation}
We note that the $(t,t)$ component of \eqref{diffeqephis} corresponds the $(x,x)$ component of Einstein's equations, and vice versa. The scalar function has the form
\begin{equation}
E_S=r^2\left(\hat\nabla^a\hat\nabla_a-\frac{2}{r}(\hat\nabla^a r)\hat\nabla_a
-\frac{(l+2)(l-1)}{r^2}\right)\phi_S+\phi_S^{(0)} \ , \label{eqephisgen}
\end{equation}
where $\phi_S^{(0)}$ is a source term determined by lower order perturbations. Using the $(t,x)$ coordinates,
\begin{equation}
E_S= \sin^2 x\left(\frac{\partial^2\phi_S}{\partial x^2}
-\frac{\partial^2\phi_S}{\partial t^2}\right)
-2\tan x\frac{\partial\phi_S}{\partial x}-(l+2)(l-1)\phi_S+\phi_S^{(0)} \ . \label{eqephistx}
\end{equation}

Viewing \eqref{diffeqephis} as a differential equation for $E_S$, the general solution is
\begin{equation}
E_S=c_1 L\tan x+\frac{L}{\cos x}\left(c_2\cos t+c_3\sin t\right) , \label{solephis}
\end{equation}
where $c_1$, $c_2$ and $c_3$ are arbitrary constants. If we change the generating function as $\phi_S\to\phi_S+\bar\phi_S$, where $\bar\phi_S$ satisfies the same differential equation as $E_S$ in \eqref{diffeqephis}, then $Z_{ab}$ remains the same in \eqref{eqzabsol}. The general form of $\bar\phi_S$ in this transformation is
\begin{equation}
\bar\phi_S=\bar c_1 L\tan x+\frac{L}{\cos x}\left(\bar c_2\cos t+\bar c_3\sin t\right) \ ,
\end{equation}
where $\bar c_1$, $\bar c_2$ and $\bar c_3$ are some other constants. Substituting into \eqref{eqephistx}, we can see that this transformation generates the same type of terms that we have in \eqref{solephis}, so by choosing $\bar c_1$, $\bar c_2$ and $\bar c_3$ appropriately, we can set $c_1=c_2=c_3=0$ in \eqref{solephis}, thereby making $E_S=0$. Hence, the $(a,b)$ components of the Einstein equations are equivalent to the single condition,
\begin{equation}
 \sin^2 x\left(\frac{\partial^2\phi_S}{\partial x^2}
 -\frac{\partial^2\phi_S}{\partial t^2}\right)
 -2\tan x\frac{\partial\phi_S}{\partial x}
 -(l+2)(l-1)\phi_S+\phi_S^{(0)}=0 \ . \label{eqphisscal}
\end{equation}

Unfortunately, it is not easy to obtain a concrete expression for the scalar source term $\phi_S^{(0)}$ in \eqref{eqphisscal} from the source terms of the $(a,b)$ components of Einstein's equations. We have to solve inhomogeneous equations of the form
\begin{equation}
\left(\hat\nabla_a\hat\nabla_b
-\frac{1}{L^2}\hat g_{ab}\right)\phi_S^{(0)}=S_{ab} \ , \label{eqphis0ab}
\end{equation}
where $S_{ab}$ are known functions of $t$ and $x$, given by lower order perturbation results. These equations can be obtained by substituting \eqref{eqephisgen} into \eqref{diffeqephis} and comparing with the $(a,b)$ components of the Einstein equations. In component form, we have to solve the following three equations for $\phi_S^{(0)}$:
\begin{align}
&\frac{\partial^2\phi_S^{(0)}}{\partial t^2}-\tan x\frac{\partial\phi_S^{(0)}}{\partial x} 
+\frac{\phi_S^{(0)}}{\cos^2 x}=S_{tt} \ , \label{eqphis0tt} \\
&\frac{\partial^2\phi_S^{(0)}}{\partial t\partial x}-\tan x\frac{\partial\phi_S^{(0)}}{\partial t}= S_{tx}\ , \label{eqphis0tx}\\
&\frac{\partial^2\phi_S^{(0)}}{\partial x^2}-\tan x\frac{\partial\phi_S^{(0)}}{\partial x} 
-\frac{\phi_S^{(0)}}{\cos^2 x}=S_{xx} \ . \label{eqphis0xx}
\end{align}
Generally, the source terms can be decomposed into a finite number of Fourier components with integer frequencies,
\begin{equation}
 S_{ab}=\sum_{\bar k=0}^{\bar k_{\mathrm{max}}}\left(S_{ab}^{(c,\bar k)}\cos(\bar k t)
 +S_{ab}^{(s,\bar k)}\sin(\bar k t)\right) ,
\end{equation}
where $S_{ab}^{(c,\bar k)}$ and $S_{ab}^{(s,\bar k)}$ are functions of $x$. We can search for the solution of \eqref{eqphis0tt}-\eqref{eqphis0xx} in a similar expansion form,
\begin{equation}
 \phi_S^{(0)}=\sum_{\bar k=0}^{\bar k_{\mathrm{max}}}\left(\phi_S^{(0,c,\bar k)}\cos(\bar k t)
 +\phi_S^{(0,s,\bar k)}\sin(\bar k t)\right) ,
\end{equation}
where $\phi_S^{(0,c,\bar k)}$ and $\phi_S^{(0,s,\bar k)}$ depend only on $x$.

When $\bar k>0$ the coefficients can be obtained in a simple algebraic way. In this case the $\cos(\bar k t)$ and $\sin(\bar k t)$ components of both \eqref{eqphis0tt} and \eqref{eqphis0tx} contain an $x$ derivative of $\phi_S^{(0,c,\bar k)}$ or $\phi_S^{(0,s,\bar k)}$. Eliminating the derivative by taking linear combinations of these two types of equations we obtain linear algebraic equations for $\phi_S^{(0,c,\bar k)}$ and $\phi_S^{(0,s,\bar k)}$, which can be solved easily.

The $\bar k=0$ case is different, since then the Fourier component of \eqref{eqphis0tx} is trivially zero. In this case \eqref{eqphis0tt} yields the differential equation
\begin{equation}
 -L\tan^2x\frac{\partial\Phi_S^{(0,c,0)}}{\partial x}=S_{tt}^{(c,0)} \quad \ , \qquad
 \Phi_S^{(0,c,0)}=\frac{1}{r}\phi_S^{(0,c,0)} \ , \label{eqkzerophi}
\end{equation}
where $r=L\tan x$, and \eqref{eqphis0xx} is identically satisfied after this. For higher orders in the $\varepsilon$ expansion it may be difficult to calculate the integral in $x$ to obtain $\Phi_S^{(0,c,0)}$. For example, at $\varepsilon^4$ order, polylogarithm functions tend to appear. Luckily, it is not necessary to compute this integral, since the calculation of the gauge invariant variables $Z_{ab}$ by \eqref{eqzabsol} using \eqref{zttphisb}-\eqref{zxxphisb} requires only the $x$ derivative of the zeroth Fourier mode of $\Phi_S=\phi_S/r$.

After the source term $\phi_S^{(0)}$ in \eqref{eqphisscal} is already determined, our last task in solving the $\varepsilon^k$ order $(l,m)$ scalar perturbation equations is to find the scalar function $\phi_S$, which corresponds to an asymptotically AdS configuration with a regular center. Using the rescaled scalar function $\Phi_S=\phi_S/r$ defined in \eqref{lowupphis}, Eq.~\eqref{eqphisscal} can be written in the simple form
\begin{equation}
-\frac{\partial^2\Phi_S}{\partial t^2}+\frac{\partial^2\Phi_S}{\partial x^2}
-\frac{l(l+1)}{\sin^2 x}\Phi_S+\frac{\Phi_S^{(0)}}{\sin^2 x}=0 \ , \label{masteqscal}
\end{equation}
where
\begin{equation}
 \Phi_S^{(0)}=\frac{1}{r}\phi_S^{(0)} \ . \label{eqbigsmallphis}
\end{equation}
Equation \eqref{masteqscal} is the master equation describing all $l\geq 2$ scalar-type perturbations.

We give a detailed description of how to solve the inhomogeneous differential equation \eqref{masteqscal} in Appendix \ref{apptimeper}. After the solution for $\Phi_S$ with the appropriate boundary condition is obtained by the method described there, then from \eqref{eqzabsol} with \eqref{zttphisb}-\eqref{zxxphisb} we obtain $Z_{ab}$, and by \eqref{zeqzab} we get $Z$. The resulting class $\mathbb{S}_{lm}$ scalar-type metric perturbation components can be obtained by \eqref{hl0hab0} and \eqref{habhaihij}.

\subsection{Case \texorpdfstring{$l=0$}{l=0}}

In this case the scalar spherical harmonic is constant, $\mathbb{S}\equiv\mathbb{S}_{00}=1/(2\sqrt{\pi})$. As it is shown in Appendix \ref{apphighordg}, for the $l=0$ scalar perturbations it is natural to make a gauge choice in which $H^{(s)}_L=0$ and $H^{(s)}_{tx}=0$ in \eqref{habhaihij}. The $(t,x)$ components of the Einstein equations gives the condition
\begin{equation}
 \frac{\partial H^{(s)}_{xx}}{\partial t}=S_{tx} \ ,
\end{equation}
where $S_{tx}$ is a function of $t$ and $x$ fixed by lower order perturbations. Integrating in $t$, we get a solution $H^{(s)}_{xx}=H_{xx}^{(s,0)}$, and we can write the general solution as
\begin{equation}
 H^{(s)}_{xx}=H_{xx}^{(s,0)}+f_1\cot x \ ,
\end{equation}
where $f_1$ is a function of $x$. Substituting into the $(t,t)$ component of the Einstein equations we get an equation determining the $x$ derivative of $f_1$. The center can be regular only if $H^{(s)}_{xx}$ has a finite limit at $x=0$, which fixes the integration constant, so we have to integrate from $x=0$, and we get a unique solution for $f_1$ and $H^{(s)}_{xx}$. 

Substituting the calculated $H^{(s)}_{xx}$ into the $(x,x)$ component of the field equations, we get an equation determining the $x$ derivative of
\begin{equation}
 \hat H^{(s)}_{tt}=H^{(s)}_{tt}\cos^2 x \ .
\end{equation}
Integrating from $x=\pi/2$, we get a solution $\hat H^{(s)}_{tt}=\hat H^{(s,0)}_{tt}$. The general solution for $H^{(s)}_{tt}$ is
\begin{equation}
 H^{(s)}_{tt}=\frac{\hat H_{tt}^{(s,0)}}{\cos^2 x}+\frac{\bar\nu}{\cos^2 x} \ , \label{eqhttnubar}
\end{equation}
where $\bar\nu$ can be any function of $t$. The $\bar\nu/\cos^2 x$ term diverges to the same order as the $g^{(0)}_{tt}$ component of the background AdS metric and corresponds to a $\varepsilon^n$ order time dependent relabeling of the $t$ coordinate, $t\to\tilde t(t)$. It is actually the gauge freedom in \eqref{eqxisttrans}. Since we are studying localized configurations, it would be unnatural to choose a time coordinate in which the metric components are asymptotically oscillating. Hence, we assume from now on that $\bar\nu$ is a constant. 

The resulting class $\mathbb{S}_{00}$ scalar-type metric perturbation components can be obtained by \eqref{habhaihij}. Since now $H^{(s)}_{tt}=2\sqrt{\pi}g_{tt}^{(k)}$, it follows that the constant $\bar\nu$ is related to $\nu_k$ defined in \eqref{nukdefeq} by
\begin{equation}
 \bar\nu=-2\sqrt{\pi}L^2\nu_k \ . \label{eqnubarnuk}
\end{equation}
Generally, we cannot set $\nu_k$ to zero, since we have assumed that the oscillation frequency of our geon is an integer in terms of the time coordinate $t$, independently of the amplitude parameter $\varepsilon$. The constant $\nu_k$ will describe the change of the physical frequency as the amplitude of the geon increases.

If there are no source terms arising from the lower order $\varepsilon$ expansion, which is certainly the case for the first order linear perturbations, then the only centrally regular solution in our gauge is $H^{(s)}_{xx}=0$ and $H^{(s)}_{tt}=\bar\nu/\cos^2 x$, which is still a gauge mode.

\section{Time-periodic solutions of the master equation} \label{apptimeper}

In this appendix we construct solutions of \eqref{eqpinhom}, which generates the time-periodic solutions of the master equation \eqref{mastereqgen}.

\subsection{Solutions of the homogeneous master equation}

For the first nontrivial order in the $\varepsilon$ expansion, i.e., for $k=1$, the source term $\Phi^{(0)}$ is necessarily zero. Even at higher orders there are no source terms for several choices of $\omega_p$. Hence we consider the $p^{(0)}=0$ case first, and we solve the homogeneous part of Eq.~\eqref{eqpinhom},
\begin{equation}
\frac{{\rm d}^2p}{{\rm d} x^2}-\frac{l(l+1)}{\sin^2 x}p+\omega_p^2 p=0 \ . \label{eqphom}
\end{equation}
Defining a new radial coordinate $z=\sin^2x$, and a rescaled function $w$ by
\begin{equation}
 p=w\sin^{l+1}x=w z^{(l+1)/2} \ ,
\end{equation}
\eqref{eqphom} can be transformed into a hypergeometric differential equation,
\begin{equation}
 z(1-z)\frac{{\rm d}^2w}{{\rm d}z^2}
 +\left[c-(a+b+1)z\right]\frac{{\rm d}w}{{\rm d}z}-abw=0 \ , \label{hipgeoeq}
\end{equation}
where the parameters are
\begin{equation}
a=\frac12(l+1-\omega_p) \ , \quad b=\frac12(l+1+\omega_p) \ , \quad c=l+\frac32 \ .
\end{equation}
A pair of fundamental solutions of the hypergeometric differential equation \eqref{hipgeoeq}, which are numerically satisfactory near the center $z=0$, are (see Sec.~15.10 of \cite{dlmf})
\begin{equation}
w_1=\,_2F_1(a,b;c;z) \quad , \quad 
w_2=z^{1-c}\,_2F_1(a-c+1,b-c+1;2-c;z) \ . \label{homsolz0}
\end{equation}
Fundamental solutions close to infinity $z=1$ are
\begin{equation}
w_3=\,_2F_1(a,b;a+b-c+1;1-z) \quad , \quad 
w_4=(1-z)^{c-a-b}\ _2F_1(c-a,c-b;c-a-b+1;1-z) \ . \label{homsolz1}
\end{equation}
The corresponding solutions of \eqref{eqphom} can be written as
\begin{equation}
 p_1=w_1\sin^{l+1}x \ , \quad
 p_2=w_2\sin^{l+1}x \ , \quad
 p_3=w_3\sin^{l+1}x \ , \quad
 p_4=w_4\sin^{l+1}x \ .
\end{equation}

Since the hypergeometric function $_2F_1(a,b;c;z)$ smoothly tends to $1$ at $z=0$, because of the $z^{1-c}$ factor, the solution $p_2$ always diverges as $x^{-l}$ and corresponds to a perturbation that is singular at the center, for both the scalar and vector cases. The solution $p_1$ always gives a regular center. The exponent in $w_4$ is $c-a-b=1/2$, and consequently $p_4$ tends to zero as $\cos x$ at infinity $x=\pi/2$, while $p_3$ tends to $1$ there. 

The transformation formula \cite{dlmf}
\begin{equation}
w_1=\frac{\Gamma(c)\Gamma(c-a-b)}{\Gamma(c-a)\Gamma(c-b)}w_3
+\frac{\Gamma(c)\Gamma(a+b-c)}{\Gamma(a)\Gamma(b)}w_4  \label{w134trans}
\end{equation}
can be used to write the centrally regular solution in terms of the fundamental solutions at infinity. Which solution generates asymptotically AdS metric perturbations at infinity depends on whether we consider scalar- or vector-type perturbations. The allowed frequencies will be different in the two cases.

In general, the function $p$ can be expanded at infinity as
\begin{equation}
 p=\sum_{j=0}^\infty \bar p^{(j)}\left(\frac{\pi}{2}-x\right)^j , \label{eqpinfexp}
\end{equation}
where $\bar p^{(j)}$ are constants. For $p=p_3$ the expansion only contains terms with even power, and $p=p_4$ only contains terms with odd $j$.

\subsection{Scalar-type linear perturbations}

For scalar-type first order linear perturbations with $l\geq 2$ substituting \eqref{eqpinfexp} and \eqref{phiexpres} into \eqref{zttphisb}-\eqref{zxxphisb}, it follows that the metric perturbation components will not have terms proportional to $\bar p^{(0)}$ that diverge as $\cos^{-2} x$. This shows that $\bar p^{(0)}$ can be arbitrary. However if $\bar p^{(1)}$ is nonzero, there will be terms diverging as $\cos^{-2} x$ both in $\hat Z_{tt}$ and in $\hat Z_{xx}$, so the metric perturbation components would not satisfy our boundary conditions \eqref{nukdefeq} and \eqref{eqgmunulim}. It is possible to show that in this case the corresponding metric is really not asymptotically AdS, by checking that the metric induced at infinity is not conformally flat. Alternatively, one can show that if $\bar p^{(1)}\not=0$ the magnetic part of the asymptotic Weyl curvature $\mathcal{B}_{ab}$ is nonzero, so the metric is not asymptotically AdS. We discuss in more detail the conditions that asymptotically AdS spacetimes have to satisfy in Appendix \ref{app:asymptotics}. 

For the solution $p=p_4$ necessarily $\bar p^{(1)}\not=0$, so the corresponding metric is not asymptotically AdS. There is no such problem with the $p_3$ solution. The perturbed metric can be asymptotically anti-de Sitter and, simultaneously, have a regular center, only if $p_1$ is proportional to $p_3$. From the transformation formula \eqref{w134trans} it follows that this is possible only if either $a$ or $b$ is a nonpositive integer. This means that asymptotically AdS scalar perturbations with a regular center only exist for frequencies $\omega_p=\omega^{(S)}_{ln}$ given by \eqref{eqscalfreq1}. In this case $a=-n$ is an integer, the hypergeometric series closes at finite order, and it can be expressed in terms of Jacobi polynomials
\begin{equation}
\,_2F_1(-n,b;c;z)=\frac{n!}{(c)_{n}}P_{n}^{(c-1,b-c-n)}(1-2z) \ , \label{jacobiident}
\end{equation}
where Pochhammer’s symbol is $(c)_{n}=\Gamma(c+n)/\Gamma(c)$. Substituting this into $p_1$, the asymptotically AdS centrally regular solution of \eqref{eqphom} for the frequency $\omega^{(S)}_{ln}$ is
\begin{equation}
p^{(S)}_{ln}=\frac{1}{L}\sin^{l+1}x\frac{n!}{(l+\frac{3}{2})_{n}}
P_{n}^{(l+\frac{1}{2},-\frac{1}{2})}(\cos(2x)) \ . \label{pslnsol}
\end{equation}
For the $ n=0$ case $w_1=1$, and the solution is simply
\begin{equation}
p^{(S)}_{l0}=\frac{1}{L}\sin^{l+1}x \ .
\end{equation}
It follows that the regular asymptotically AdS scalar-type solutions of the homogeneous part of \eqref{masteqscal} with the given frequency can be written as in \eqref{scallinpert1} and \eqref{scallinpert2}.

\subsection{Vector-type linear perturbations}

For $l\geq 2$ vector-type linear perturbations, it follows from \eqref{eqpinfexp} and \eqref{phiexpres} that the expansion of $\phi_V=r\Phi_V$ starts as
\begin{equation}
 \phi_V=\left[\frac{\bar p^{(0)}}{\frac{\pi}{2}-x}+\bar p^{(1)}
 +\left(\bar p^{(2)}-\frac{\bar p^{(0)}}{3}\right)\left(\frac{\pi}{2}-x\right)
 +\cdots\right]L\cos(\omega_p t) \ ,
\end{equation}
or a similar expression with $\cos(\omega_p t)$ replaced by $\sin(\omega_p t)$. Using \eqref{eqsztzx} it follows that if $\bar p^{(0)}\not=0$ then $Z_t$ diverges as $\cos^{-2} x$, and from \eqref{eqzhgauge} and \eqref{vectgauge} it can be seen that the metric perturbation components $g^{(k)}_{ti}$ also diverge as $\cos^{-2} x$, so the metric is not asymptotically AdS.

The solution $p=p_3$ has a nonzero $\bar p^{(0)}$, so it cannot belong to an asymptotically AdS metric perturbation. The metric can be asymptotically anti-de Sitter with a regular center only if $p_1$ is proportional to $p_4$. Since by \eqref{w134trans} this is possible only if $c-a$ or $c-b$ is a nonpositive integer, it follows that regular vector-type perturbations only exist for frequencies given in \eqref{eqvectfreq1}. In this case $c-b=-n$, and the series in $w_4$ closes at finite order. Using \eqref{w134trans}, \eqref{jacobiident}, and the symmetry property
$P_{n}^{(\alpha,\beta)}(-z)=(-1)^{n} P_{n}^{(\beta,\alpha)}(z)$,
the regular asymptotically AdS solution of \eqref{eqphom} turns out to be
\begin{equation}
p^{(V)}_{ln}=\frac{1}{L}\sin^{l+1}x\cos x\frac{n!}{(l+\frac{3}{2})_{n}}
P_{n}^{(l+\frac{1}{2},\frac{1}{2})}(\cos(2x)) \ . \label{eqpvlndef}
\end{equation}
The $n=0$ solution is
\begin{equation}
p^{(V)}_{l0}=\frac{1}{L}\sin^{l+1}x\cos x \ .
\end{equation}
The regular asymptotically AdS vector generating function that solves the homogeneous part of \eqref{eqmastvect} with the given frequency is given in \eqref{vectlinpert1} and \eqref{vectlinpert2}.

\subsection{Solutions of the inhomogeneous master equation} \label{subsecperinhom}

Let us consider two linearly independent solutions of the homogeneous equation \eqref{eqphom}. For the first solution we take the previously defined $p_1$, which is always regular at the center. We denote the yet unspecified second solution by $p_b$, and we assume that it corresponds to an asymptotically AdS linear metric perturbation (scalar- or vector-type). Then the regular asymptotically AdS solution of the inhomogeneous equation \eqref{eqpinhom} can be obtained as
\begin{equation}
 p=p_1\int_{\frac{\pi}{2}}^x\frac{p_b p^{(0)}}{W\sin^2 x}{\rm d}x
 -p_b\int_0^x\frac{p_1 p^{(0)}}{W\sin^2 x}{\rm d}x \ , \label{eqpintsol}
\end{equation}
where the Wronskian is
\begin{equation}
 W=p_1\frac{{\rm d}p_b}{{\rm d}x}-p_b\frac{{\rm d}p_1}{{\rm d}x} \ .
\end{equation}
Since there is no first derivative term in \eqref{eqpinhom}, by Abel's differential equation identity follows that the Wronskian of any two solutions of the homogeneous equation is a constant. This constant is not zero if and only if the two solutions are linearly independent. In the solution \eqref{eqpintsol}, the singularity of $p_b$ at the center is compensated by the choice of $x=0$ as the lower limit in the second integral. Similarly, the choice of $\pi/2$ as the lower limit of the first integral is to compensate for the behavior of $p_1$ at infinity.

For scalar-type perturbations the asymptotically AdS solution is obtained by setting $p_b=p_3$, while for vector perturbations we have to choose $p_b=p_4$. The above argument, that \eqref{eqpintsol} gives indeed the asymptotically AdS metric, takes into account only the contribution from the $\varepsilon^k$ order homogeneous terms. However, in each concrete case it is easy to check that the inhomogeneous terms do not give any terms in the metric perturbation components $g^{(k)}_{\mu\nu}$ that diverge as $\cos^{-2} x$, so our boundary conditions \eqref{nukdefeq} and \eqref{eqgmunulim} are satisfied, and the generated metric is really asymptotically AdS.

In certain resonant cases, the solutions $p_1$ and $p_b$ chosen in the above way are not linearly independent. In these cases $p_1$ is already behaving well both at the center and at infinity. This happens exactly when the frequency $\omega_p$ of the source term agrees with the resonant frequencies given by \eqref{eqscalfreq1} or \eqref{eqvectfreq1} belonging to the actual value of $l$ for some non-negative integer $n$. In these resonant cases the only available second solution is $p_b=p_2$, which is singular both at $x=0$ and at $x=\pi/2$. One  solution of the inhomogeneous equation can still be calculated by \eqref{eqpintsol}. However, the singularity of $p_b=p_2$ both at the center and at infinity can only be compensated in the second term of \eqref{eqpintsol} if we require the condition \eqref{eqrescond} given in the main text. The factor $W$ is dropped from the denominator since it is a nonzero constant.

Equation \eqref{eqrescond} is a very important consistency condition on the source term $p^{(0)}$ in the resonant cases. If this conditions fails to hold, then there are no time-periodic centrally regular asymptotically AdS perturbations of the system. Even if this condition fails then there are centrally regular asymptotically AdS perturbations, but they must be secularly growing with time dependence of the type $t\cos(\omega_p t)$ or $t\sin(\omega_p t)$. However, that type of time dependence is excluded by the formalism applied in this paper, since we are only working with periodic solutions now.

For the resonant case the lower limit in the first integral of \eqref{eqpintsol} can be left arbitrary, since $p_1$ is behaving well at infinity. This corresponds to the freedom of adding the homogeneous solution $p_1$ multiplied by an arbitrary constant. In these resonant cases the centrally regular asymptotically AdS solutions of the inhomogeneous equation are
\begin{equation}
 p=p_1\int_{\frac{\pi}{2}}^x\frac{p_2 p^{(0)}}{W\sin^2 x}{\rm d}x
 -p_2\int_0^x\frac{p_1 p^{(0)}}{W\sin^2 x}{\rm d}x + c_p p_1 \ , \label{eqpintinhom}
\end{equation}
where $c_p$ is an arbitrary constant, and we have to require \eqref{eqrescond}. In many cases, the condition \eqref{eqrescond} can be used to fix the value of some other constant $\bar c_p$, which had to be introduced earlier at a lower order in the $\varepsilon$ expansion.

\subsection{Static mode of scalar perturbations}

The general method in the previous subsection can also be applied to the zero frequency part of the source function $\phi_S^{(0)}$, when $\omega_p=0$ in \eqref{eqpinhom}. However, in this case there is a computationally less involved method to obtain the corresponding metric perturbation. We have noted in connection to \eqref{eqkzerophi} that for the zero frequency case it is easy to obtain the $x$ derivative of the source function $p^{(0)}$, but it can be difficult to perform the integral to get the actual form of $p^{(0)}$. Luckily, for the static part of the metric perturbation we only need the derivative of the function $p$, which can be seen from \eqref{zttphisb}-\eqref{zxxphisb}. If $\omega_p=0$, then we can multiply \eqref{eqpinhom} by $\sin^2x$, and after taking the derivative we can write the equation into the form
\begin{equation}
\frac{{\rm d}^2\bar p}{{\rm d}x^2}-\frac{l(l+1)}{\sin^2 x}\bar p+\bar\omega_{\bar p}^2\bar p
+\frac{\bar p^{(0)}}{\sin^2 x}=0 \ ,
\end{equation}
where
\begin{equation}
 \bar p=\sin x\frac{{\rm d}p}{{\rm d}x} \quad , \qquad
 \bar p^{(0)}=\sin x\frac{{\rm d}p^{(0)}}{{\rm d}x} \quad , \qquad
 \bar\omega_{\bar p}=1 \ .
\end{equation}
This equation can be solved by the method described in Sec.~\ref{subsecperinhom}, with the only difference that now the asymptotically AdS solutions belong to $p_4$ for scalar-type perturbations. After getting the solution for $\bar p$, the metric perturbations can be calculated using \eqref{zttphisb}-\eqref{zxxphisb}, without calculating $p$.

\subsection{Static mode of vector perturbations}

The time-independent part of vector-type perturbations can also be calculated by the methods described in Sec.~\ref{subsecperinhom}. However, it is also possible to proceed without computing the integral that is necessary to obtain the function $\phi_V^{(0,c,0)}$ in the expansion \eqref{eqphiv0exp}. We have to solve the static part of the equation $E_V=0$, where $E_V$ is given by \eqref{evexpression}. We substitute $\phi_V=q$ and $\phi_V^{(0)}=q^{(0)}$, where $q$ and $q^{(0)}\equiv\phi_V^{(0,c,0)}$ are functions of only $x$,
\begin{equation}
 \sin^2 x\frac{{\rm d}^2q}{{\rm d}x^2}-2\tan x\frac{{\rm d}q}{{\rm d}x}
 -(l+2)(l-1)q+q^{(0)}=0 \ .
\end{equation}
Taking the derivative of the equation and introducing
\begin{equation}
 \bar q=\frac{{\rm d}q}{{\rm d}x} \quad , \qquad
 \bar q^{(0)}=\frac{{\rm d}q^{(0)}}{{\rm d}x}  \ ,
\end{equation}
we obtain
\begin{equation}
 \sin^2 x\frac{{\rm d}^2\bar q}{{\rm d}x^2}
 -2\frac{\sin^3 x}{\cos x}\frac{{\rm d}\bar q}{{\rm d}x}
 -\frac{2}{\cos^2x}\bar q-(l+2)(l-1)\bar q+\bar q^{(0)}=0 \ .
\end{equation}
We know the source term $\bar q^{(0)}$, and to get the metric perturbation by \eqref{eqsztzx} we only need $\bar q$. This equation can be solved in a similar way as we have solved \eqref{eqpinhom}. The four fundamental solutions $q_\alpha$ for $\alpha=1,2,3,4$ can be given by  $q_\alpha=w_\alpha\sin^{l+1}x\cos x$, with $w_\alpha$ given in \eqref{homsolz0} and \eqref{homsolz1}, where now
\begin{equation}
 a=\frac{l}{2}+1 \ , \quad b=\frac{l}{2}+2 \ , \quad c=l+\frac32 \ .
\end{equation}
The inhomogeneous solution for $\bar q$ can be obtained as in \eqref{eqpintsol}, with the replacements $p\to\bar q$, $p^{(0)}\to\bar q^{(0)}$, $p_1\to q_1$, and $p_b\to q_3$. When $\bar q$ is known, it determines the $x$ derivative of the static part of $\phi_V$, and the metric perturbation variables can be calculated by \eqref{eqsztzx}, \eqref{eqzhgauge}, and \eqref{vectgauge}.

\bibliography{omthree}

\end{document}